\documentclass[a4paper,fleqn,usenatbib]{mnras}

\usepackage[T1]{fontenc}
\usepackage{ae,aecompl}

\usepackage{graphicx}	
\usepackage{amsmath}	
\usepackage{amssymb}	

\usepackage{natbib}
\usepackage{hyperref}
\usepackage{multirow}
\usepackage{lscape}
\usepackage[percent]{overpic}
\usepackage{float}
\usepackage{xcolor}
\usepackage{pict2e}
\usepackage[font=footnotesize]{subfig}
 \usepackage[flushleft]{threeparttable}

\title[Bar--driven secular evolution in the $\gamma-\sy$ galaxy \so]{Evidence of bar--driven secular evolution in the gamma--ray narrow--line Seyfert 1 galaxy FBQS J164442.5+261913}
 
\author[A. Olgu\'in--Iglesias]{
A. Olgu\'in--Iglesias,$^{1,2}$\thanks{E-mail: olguin@inaoep.mx}
J. K. Kotilainen,$^{2}$
J. Le\'on Tavares,$^{3}$
V. Chavushyan$^{1}$
C. A\~norve$^{4}$
\\
\author[A. Olgu\'in--Iglesias et al.]{}
\\
$^{1}$Instituto Nacional de Astrof\'isica \'Optica y Electr\'onica (INAOE), Apartado Postal 51 y 216, 72000 Puebla, M\'exico;\\
$^{2}$Finnish Centre for Astronomy with ESO (FINCA), University of Turku, V\"ais\"al\"antie 20, FI-21500 Piikki\"o, Finland\\
$^{3}$Sterrenkundig Observatorium, Universiteit Gent, Krijgslaan 281-S9, B-9000 Gent, Belgium\\
$^{4}$Facultad de Ciencias de la Tierra y del Espacio de la Universidad Aut\'onoma de Sinaloa, Blvd. de las Am\'ericas y Av.\\
Universitarios S/N, Ciudad Universitaria, C.P. 80010, Culiac\'an
Sinaloa, M\'exico\\
}

\date{Accepted XXX. Received YYY; in original form ZZZ}

\pubyear{2015}

\begin{document}
\label{firstpage}
\pagerange{\pageref{firstpage}--\pageref{lastpage}}
\maketitle

\newcommand{\so}{FBQS J164442.5+261913}
\newcommand{\sy}{NLSy1}
\newcommand{\g}{$\gamma$}

\begin{abstract} 
We present near--infrared (NIR) imaging  of \so, one of the few $\gamma$--ray emitting Narrow Line Seyfert 1 (\sy) galaxies detected at high significance level by $Fermi$--LAT. This study is the first morphological analysis performed of this source and the third performed of this class of objects. Conducting a detailed two--dimensional modeling of its surface brightness distribution and analysing its $J-K_s$ colour gradients, we find that \so\ is statistically most likely hosted by a barred lenticular galaxy (SB0). We find evidence that the bulge in the host galaxy of \so\ is not classical but pseudo, against the paradigm of powerful relativistic jets exclusively launched by giant ellipticals. Our analysis, also reveal the presence of a ring with diameter equalling the bar length ($r_{bar} = 8.13\ \textrm{kpc}\pm 0.25$), whose origin might be a combination of bar--driven gas rearrangement and minor mergers, as revealed by the apparent merger remnant in the $J$--band image. 
In general, our results suggest that the prominent bar in the host galaxy of \so\ has mostly contributed to its overall morphology driving a strong secular evolution, which plays a crucial role in the onset of the nuclear activity and the growth of the massive bulge. Minor mergers, in conjunction, are likely to provide the necessary fresh supply of gas to the central regions of the host galaxy. \smallskip

\end{abstract}



\section{Introduction}\label{sec:intro}

Narrow line Seyfert 1 (\sy) galaxies are type 1 active galactic nuclei (AGN) characterized by narrower Balmer lines ($FWHM(H\beta)<2000\ km\ s^{-1}$) than in normal Seyferts, flux ratios $[OIII]/ H\beta <3$, strong optical FeII lines (FeII bump) and a soft X-ray excess \citep{osterbrok_pogge_1985,pogge_2000}. Based on the full width at half maximum (FWHM) of their Broad Line Region (BLR) lines and the continuum luminosity \citep{kaspi_2000}, their central black holes masses ($M_{BH}$) are estimated to range from $\sim 10^6M\odot$ to $\sim10^7M\odot$ \citep[][although \citealt{baldi_2016} show that these low $M_{BH}$ estimates might be seriously affected by the orientation of the BLR]{mathur_2012}. Their low--mass black holes suggest that their accretion rates are close to the Eddington limit and their host galaxies are in an early phase of galaxy evolution \citep{ohta_2007}. Unfortunately, relatively little is known about their host galaxies. 

Some studies find that their morphologies resemble those of inactive spirals with a regular presence of stellar bars \citep{crenshaw_2003}, and pseudobulges \citep{orban_2011,marthur_2012}. However, \g--ray\ emission have been detected in seven radio--loud \sy\ (RL--\sy) by the Large Area Telescope (LAT) on board the \textit{Fermi} satellite, suggesting that highly beamed and strongly collimated relativistic jets can be launched by RL--\sy\ AGN. The latter, challenge the paradigm that such jets are launched exclusively by blazars hosted by giant elliptical galaxies \citep{laor_2000,marscher_2009} with black holes with masses $ M_{BH}\gtrsim10^8M_{\odot} $ accreting at low rates \citep{mclure_2004,sikora_2007}. Therefore, a thorough analysis of the host galaxies of this new class of AGN \citep[hereafter \g--\sy,][]{fermi_nlsy1}, becomes a priority.\smallskip

So far, only two \g-- \sy\ host galaxies have been characterized, 1H 0323+342 \citep{anton_2008,leontavares_2014} and PKS 2004-447 \citep{kotilainen_2016}. These studies reveal characteristics such as the presence of disks, rings, bars and pseudobulges, which are expected in normal \sy s, however, do not fit with the common belief that powerful relativistic jets are launched exclusively by giant ellipticals.\smallskip

As part of our ongoing imaging survey of the complete sample of \g--\sy\ galaxies detected so far, we conducted NIR ($J$ and $K_s$--bands) observations to the \g--\sy\ \so. This is one of the sources detected by \textit{Fermi}--LAT with high significance, having test statistic $TS>25$ \citep[$\sim 5\sigma$,][]{mattox_1996} and given its redshift \citep[$z=0.145$,][]{bade_1995}, it is the second closest after 1H 0323$+$342 ($z=0.061$), making it an excellent candidate for accurate morphological studies to its host galaxy. With the aim of achieving a better understanding of the mechanisms needed to form and develop highly collimated relativistic jets, in this paper we present the results from our thorough analysis to \so. \smallskip

This paper is structured as follows: Observations and data reduction are presented in Section 2; the methods we adopt to analyse the data are explained in Section 3. Our results and discussion are presented in Section 4 and 5. In Section 6 we summarize our findings. Throughout the manuscript we adopt a concordance cosmology with $\Omega_{m}=0.3$, $\Omega_{\Lambda} = 0.7$ and a Hubble constant of   $H_{0} = 70$ Mpc$^{-1}$ km s$^{-1}$. 


\section{Observations and data reduction}\label{sec:images}

The $J$-- and $K$--band observations of \so\ were conducted at the 2.5 m Nordic Optical Telescope (NOT) during the night of May 1, 2015 using the Wide--Field near--infrared camera NOTcam with CCD dimensions of 1024 pix $ \times $ 1024 pix and a pixel scale of $0.234''/ \textrm{pix}$, giving a field of view of $\sim4\times4\ \textrm{arcmin}^2 $). During the night, the seeing was very good, with an average FWHM of $\sim0.75''$ and $\sim0.63''$ for $J$-- and $K_s$--bands, respectively. The target was imaged using the NOTcam standard $J$ ($\lambda_{central}=1.246\mu m$) and $K_s$ ($\lambda_{central}=2.140\mu m $) filters with a dithering technique with individual exposures of 30 seconds and a typical offset of $\sim10''$. A total of 85 individual exposures for $J-$band and 72 for $K_s-$band were obtained, giving a total exposure time of 2550 seconds and 2160 seconds, respectively.\smallskip

The data reduction was performed using the NOTcam reduction package\footnote{\url{www.not.iac.es/instruments/notcam/}} for IRAF\footnote{IRAF   is   distributed   by   the   National   Optical   Astronomy Observatories, which are operated by the Association of Universities for Research in Astronomy,  Inc., under  cooperative  agreement  with the National Science Foundation}. First we corrected for the optical distortion of the Wide--Field camera using distortion models based on high quality data of a stellar--rich field. Then, bad pixels were masked out using a file available in the NOTCam bad pixel mask archive. A normalized flat field was created from evening and morning sky frames to account for the thermal contribution. Using field stars as reference points, the dittered images were aligned and co-added to obtain the final reduced image used in our analysis. In order to perform photometric calibration to the images, we retrieved $J$-- and $K_s$-- band magnitudes from 2MASS \citep[]{2mass} resulting in an accuracy of $ \sim0.10\ \textrm{mag} $. The derived integrated magnitudes in circular apertures are $m_J=15.35\pm0.10$ ($M_J=-23.84\pm0.10$) and $m_{Ks}=13.44\pm0.10$ ($M_{Ks}=-25.86\pm0.10$). Galactic extinction for $J$ and $K_{s}$ bands are negligible ($A_\lambda [J]=0.058$ and $A_\lambda [K_s]=0.025$).\smallskip 

\section{Image analysis}
\subsection{Photometric decomposition}
We perform a 2D modeling of the galaxy using the image decomposition code GALFIT \citep{peng_2011}. We follow the procedure described in our previous studies of AGN host galaxies \citep{leontavares_2014, olguin-iglesias_2016}, which is described below.\smallskip

\begin{figure*}
\begin{overpic}[width=\textwidth]{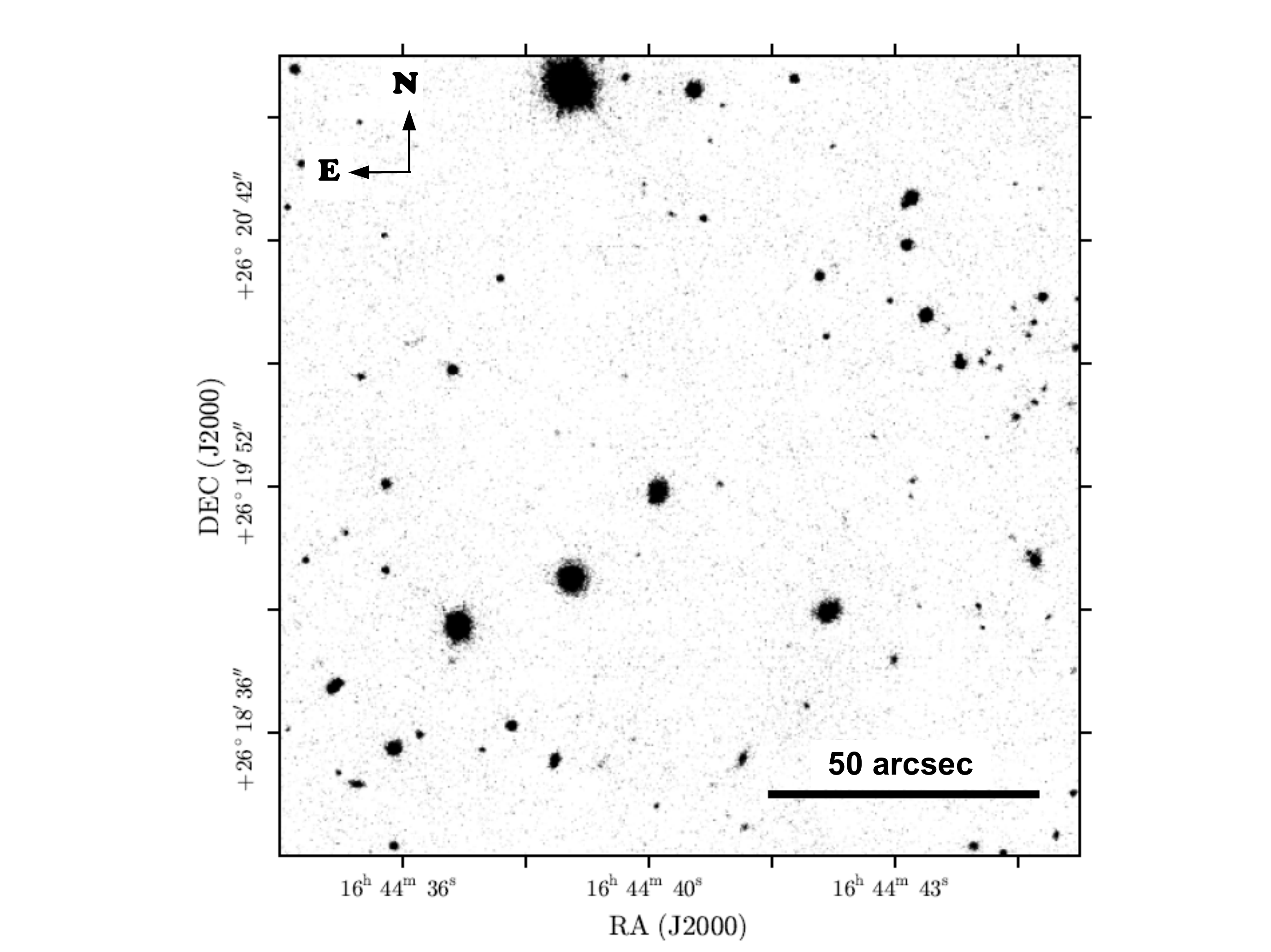}
\linethickness{2pt}
\put(38,68.3){\color{red}\vector(1.,0){3.5}} 
 \put (36.2,67.4) {\huge 1}
\put(59.,67.8){\color{red}\vector(-1.,0){3.5}} 
\put(59.5,66.7){\huge3} 
\put(67.5,59.3){\color{red}\vector(1.,0){3.5}} 
\put(65.5,58.7){\huge4} 
\put(71.3,46.3){\color{red}\vector(1.,0){3.5}} 
\put(69.3,45.5){\huge7} 
\put(77.8,51.5){\color{red}\vector(1.,0){3.5}} 
\put(74.5,50.7){\huge10} 
\linethickness{3pt}
\put(29.8,25.6){\color{blue}\vector(3.5,0){5.}} 
\put(27.5,24.8){\huge 2} 
\put(67.,50.1){\color{blue}\vector(3.5,0){5}} 
\put(65.,49.3){\huge5} 
\put(65.5,55.6){\color{blue}\vector(3.5,0){5}} 
\put(63.5,54.8){\huge 6} 
\put(46.,17.8){\color{blue}\vector(-3.5,0){5}} 
\put(47.,16.9){\huge8} 
\put(59.6,48.5){\color{blue}\vector(3.5,0){5}} 
\put(57.6,47.8){\huge9} 
\linethickness{3pt}
\put(44.9,38.){\color{green}\vector(0,-7.5){7.}} 
\put(33.,38.9){\large\so} 
\end{overpic}
\caption{$J$--band NOTcam image of \so. The large green vertical arrow indicates the location of the target. Horizontal arrows show the suitable (blue thick arrows) and unsuitable (red thin arrows) stars for the PSF construction.}
\label{fig:FOV}
\end{figure*}

The first, and most important part of the analysis is the modeling of the point spread function (PSF) by fitting selected stars of the field of view (FOV, Fig. \ref{fig:FOV}). These stars are non-saturated, with no sources within $\sim7''$ radius, more than $\sim10''$ away from the border of the FOV and with a range of magnitudes that allow us a proper characterization of core and wings. Stars 2, 5, 6, 8 and 9 fulfil these criteria (see Figure \ref{fig:FOV}) and thus are used to derive our PSF model. On the other hand, star 1 is saturated, stars 3 and 10 are very close to the border of the FOV and stars 4 and 7 have close companions. 

\begin{figure*}
\centering
\begin{tabular}{c c }
\hspace*{0.6cm}\includegraphics[clip, trim=0.5cm -3.cm 0.5cm 0.0cm,width=0.38\textwidth]{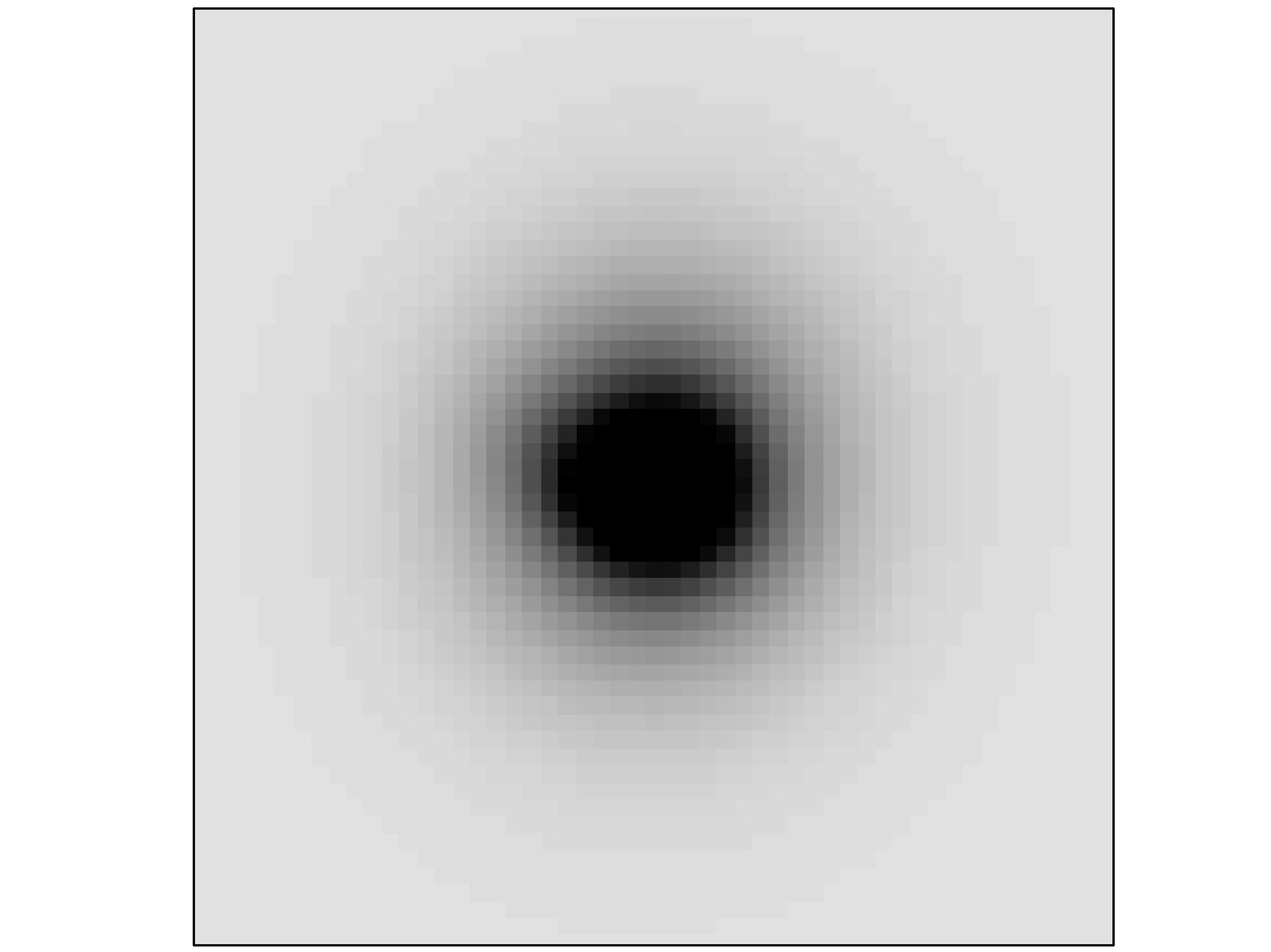} &
\includegraphics[clip, trim=0.5cm 0.4cm 0.7cm 0.7cm,width=0.35\textwidth]{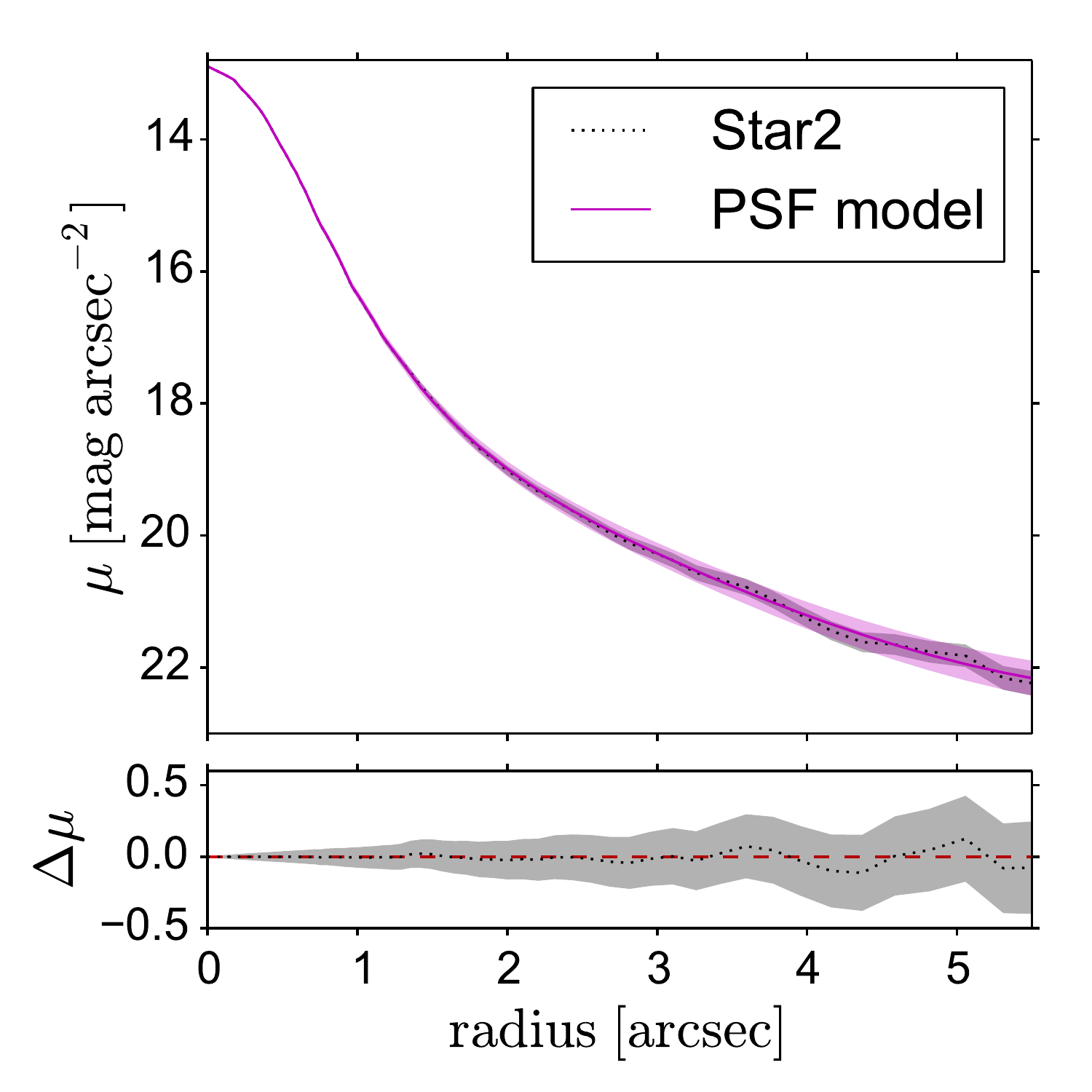} \\
\includegraphics[clip, trim=0.5cm 0.4cm 0.7cm 0.8cm,width=0.35\textwidth]{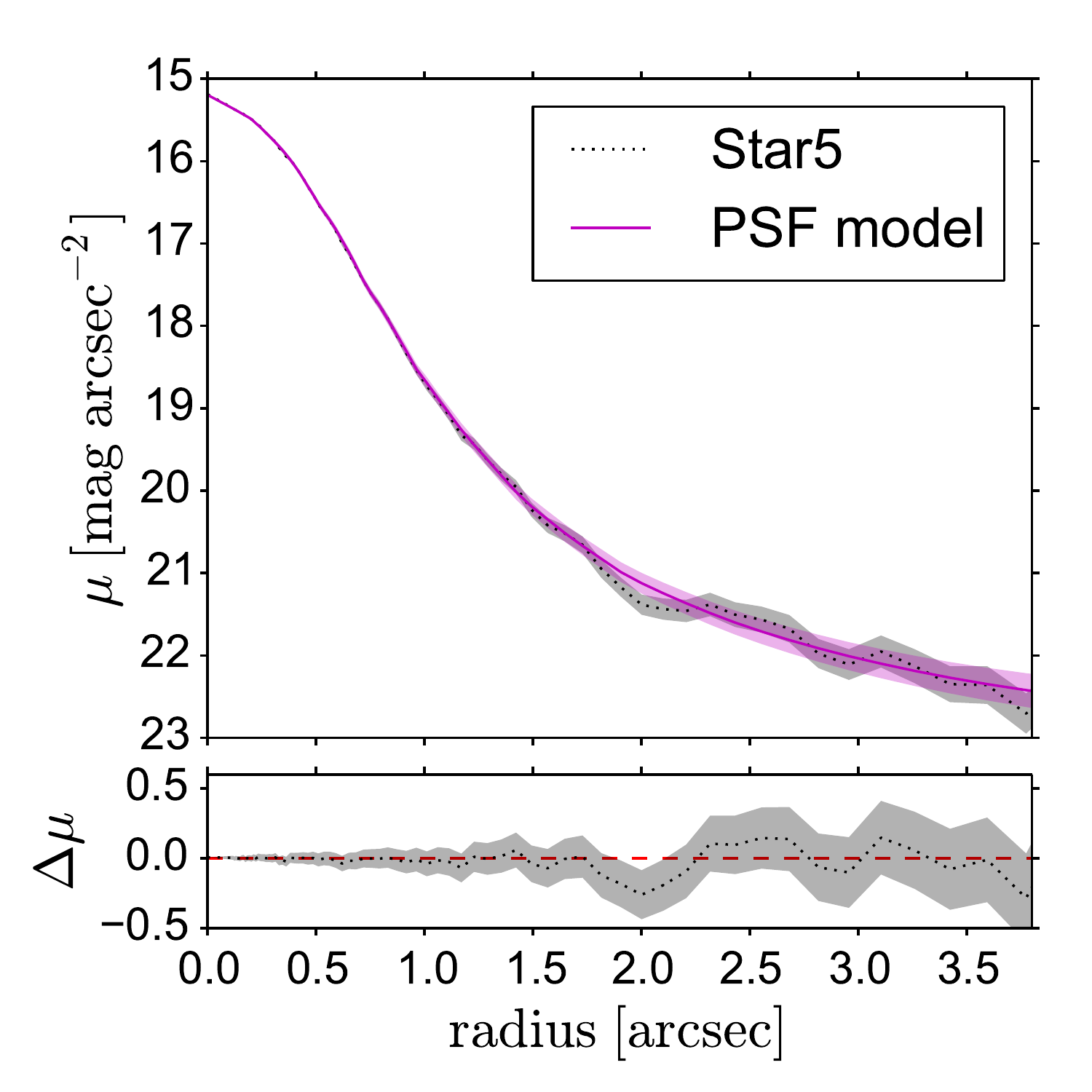} &
\includegraphics[clip, trim=0.5cm 0.4cm 0.7cm 0.8cm,width=0.35\textwidth]{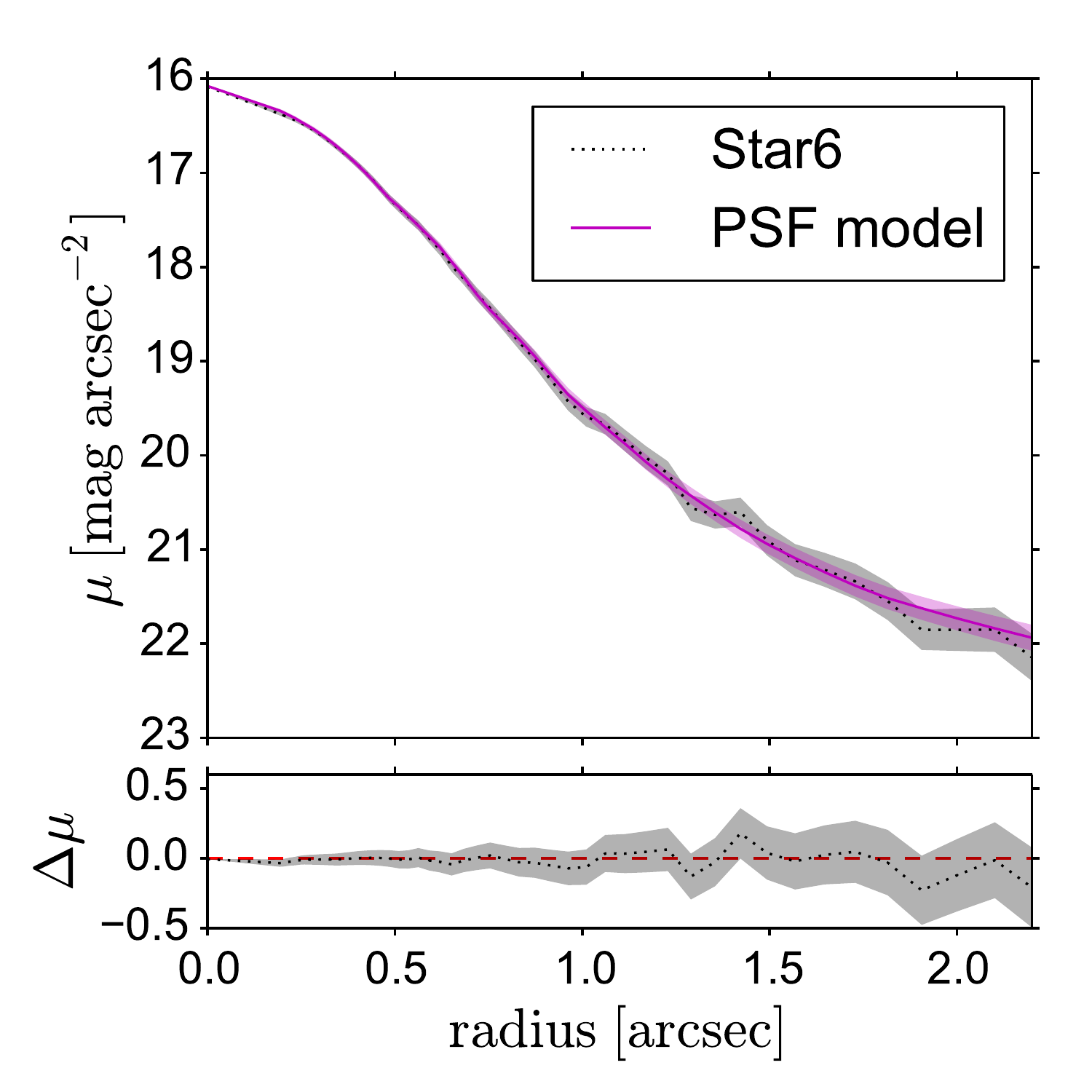} \\
\includegraphics[clip, trim=0.5cm 0.4cm 0.7cm 0.8cm,width=0.35\textwidth]{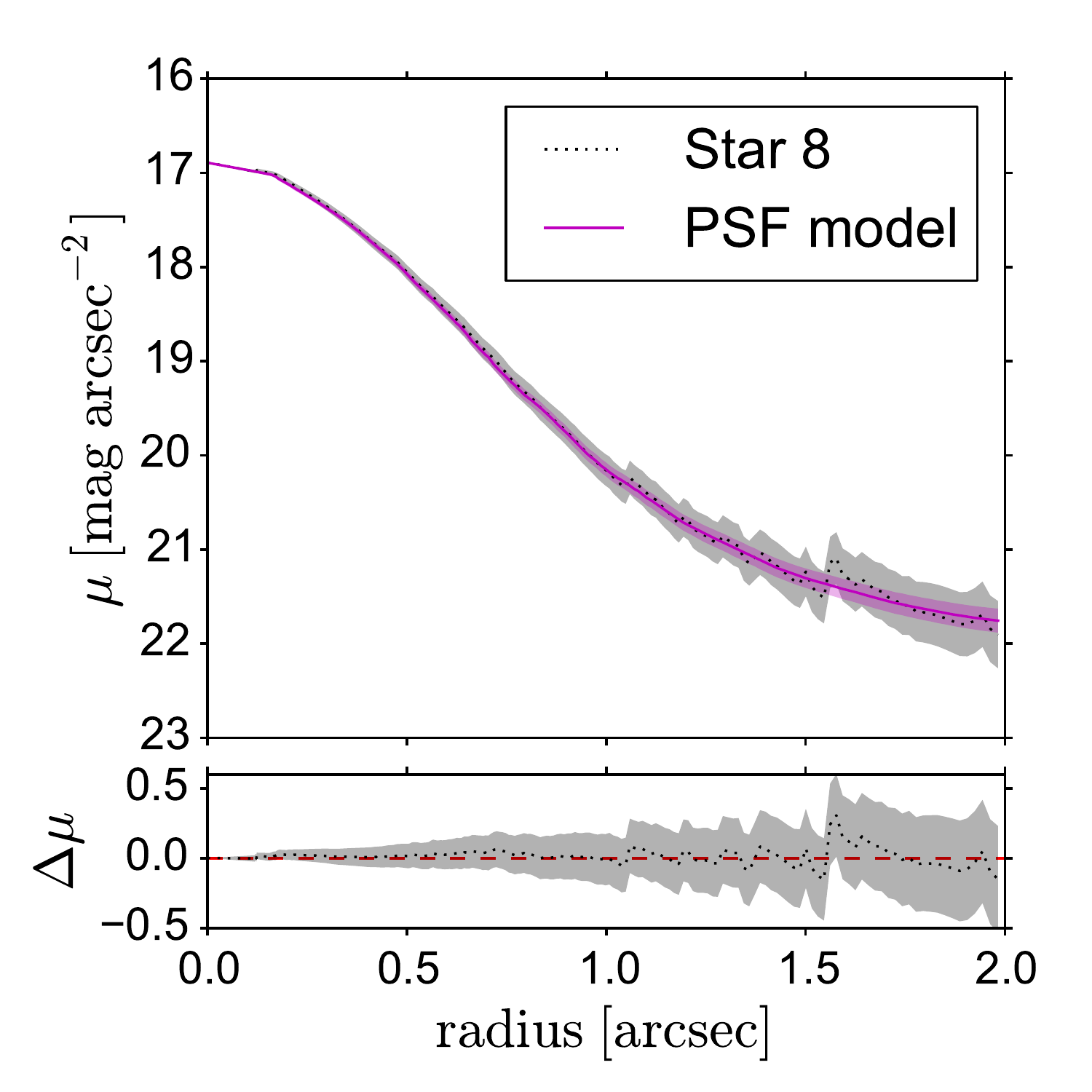} &
\includegraphics[clip, trim=0.5cm 0.4cm 0.7cm 0.8cm,width=0.35\textwidth]{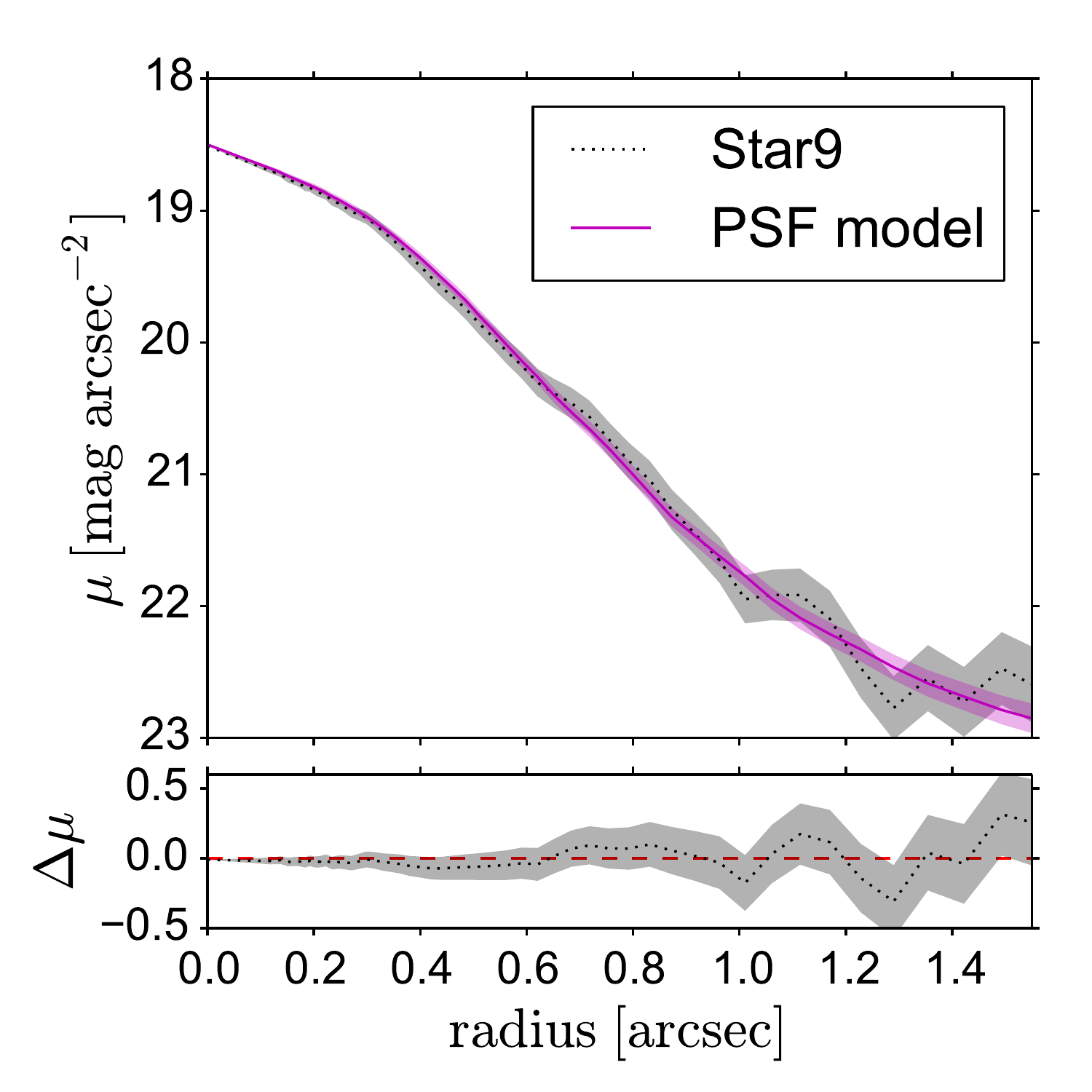} \\
   \end{tabular}
  \caption{Test of the PSF model (top left image). Each star is fitted with our final PSF model in order to ensure its reliability. The top subpanel of each plot shows the azimuthally averaged surface brightness profiles of the PSF model (magenta line) and the fitted star (black data points). The lower subpanel shows the residuals of the fit.}
  \label{psf_test}
\end{figure*}

Each selected star is centered in a $50''\times50''$ box, where all extra sources are masked out by implementing the segmentation image process of SExtractor \citep{bertin_1996}. The stars are simultaneously modeled, using one Gaussian function (intended to fit the core of the stars) then, the resulting model is added with an exponential function (intended to fit the wings of the stars). Similarly, depending on the residuals, we add extra Gaussians and exponential functions until the core and wings are satisfactorely fitted. For our imagery, six Gaussians and six exponentials (and a flat plane, that fits the sky background) were enough. The result is considered as a suitable PSF model for our analysis once it successfully fits all the stars individually (Figure \ref{psf_test}).\smallskip

Next, we fit \so\ with the scaled version of our derived PSF as the only component, to constrain the unresolved AGN contribution at the centre of the galaxy. Since the residuals of the single PSF model (hereafter model 1) are considerable ($\chi^2_{model 1}=4.148\pm0.01$ for $J$--band and $\chi^2_{model 1}=3.171\pm0.02$ for $K_s$--band), we continue our analysis by adding extra functions to the model. We use the S\'ersic profile, expressed such that

    \begin{center}
	\begin{equation}
	I(R)=I_e exp\left[ -\kappa_n\left( \left(\frac{R}{R_e} \right)^{1/n} -1\right)\right]
	\label{sersic}
	\end{equation}
	 \end{center}

\noindent
where $I(R)$ is the surface brightness at the radius R, and $\kappa_n$ is a parameter coupled to the S\'ersic index $n$ in such a way that $I_e$ is the surface brightness at the effective radius $R_e$, where the galaxy contains half of the light \citep{Graham_2005}. The S\'ersic profile has the ability to represent different stellar distributions such as elliptical galaxies, classical-- and pseudo--bulges and bars, just by varying its S\'ersic index $n$. Hence, when $n = 4$, the S\'ersic funtion is known as the de Vaucouleurs profile (widely used to fit elliptical galaxies and classical bulges); when $n = 1$, it is an exponential function, and when $n = 0.5$, it is a Gaussian.\smallskip

Given that \sy s are known to be typically hosted in disc galaxies \citep{creenshaw_2003}, we also explore models that include the exponential function, expressed as:

    \begin{center}
	\begin{equation}
	I(R)=I_0 exp  \left(\frac{R}{h_r} \right)
	\label{eq:sersic}
	\end{equation}
	 \end{center}
\noindent
where $I(R)$ is the surface brightness at the radius $R$, $I_0$ is the central surface brightness and $h_r$ is the disc scale length.\smallskip

\subsubsection{Uncertainties} \label{errors_section}
Since the error bars produced by GALFIT are purely statistical and thus, unrealistically small \citep{haussler_2007, bruce_2012}, we follow \cite{kotilainen_2011} and \cite{leontavares_2014} to derive the uncertainties of our fittings. We identify model parameters that could contribute most significantly to errors. Regarding the PSF, spatial variations might affect the structural parameters of the galaxy model and, to a lesser extent, its magnitudes. To account for this, we compare the brightness distribution of our PSF model with the brightness distribution of each star in the field, whose only difference is assumed to lie in their positions.\smallskip 

On the other hand, sky background can affect magnitudes in a larger extent (when compared to the PSF) and to a lesser extent (yet significantly), the structural parameters of the galaxy model. Even though, our imagery is in NIR bands and thus the sky counts are $\textrm{SKYCOUNTS}\approx0$, they may show large variations. To account for this, we run several sky fits in separated regions of 300 pixels $\times$ 300 pixels ($70''\times70''$) and use the mean and $\pm1\sigma$ of the resultant values to fit the galaxy. The outcomes are models with slightly different magnitudes which are assumed to be the errors due to the sky background.\smallskip

\begin{figure*}
\centering  
\begin{tabular}{rr}
\includegraphics[clip, trim=2.5cm 1.5cm 1.5cm .4cm,width=0.35\textwidth]{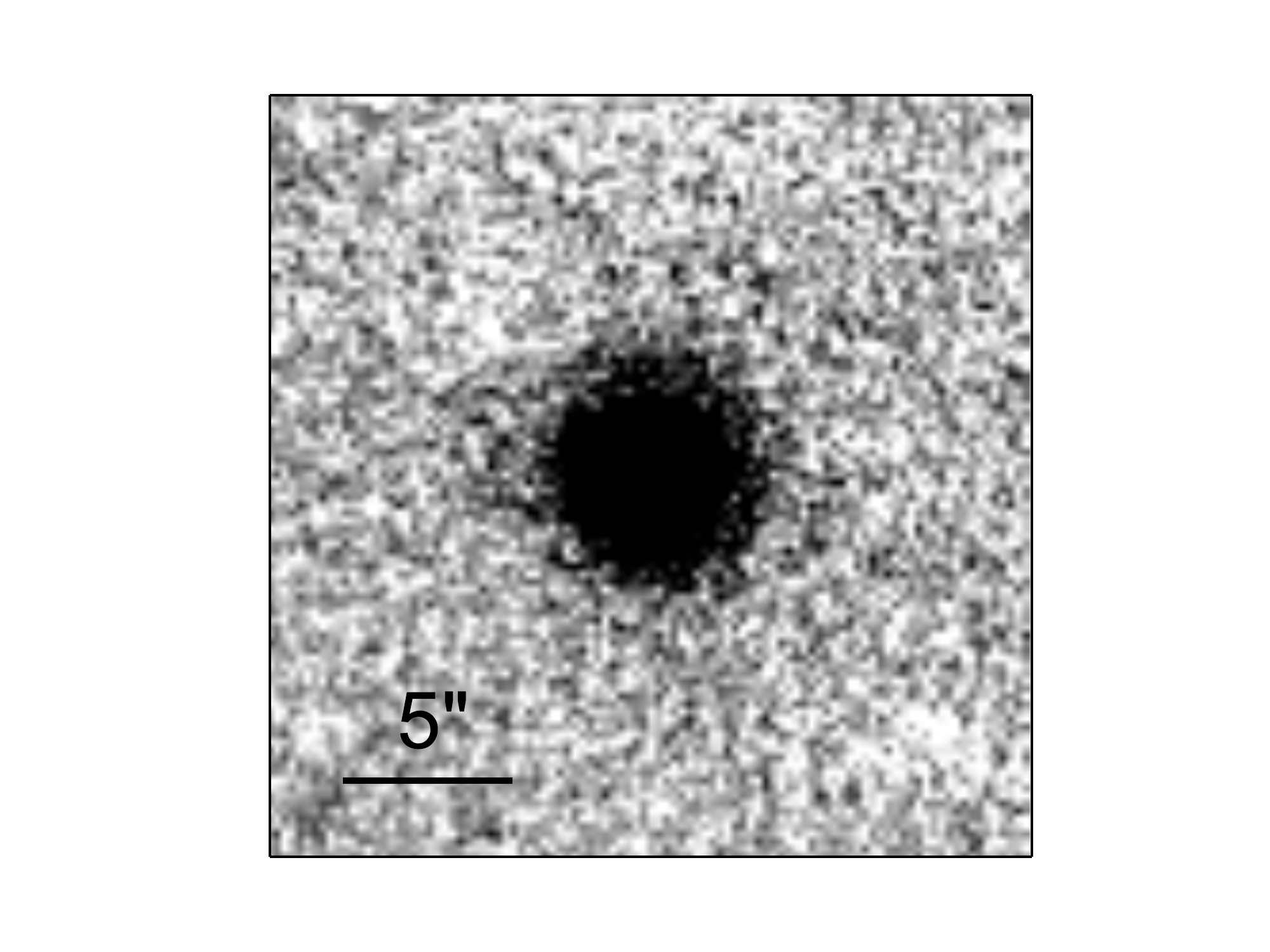} & \includegraphics[clip, trim=1.5cm 1.5cm 2.5cm 1.4cm,width=0.35\textwidth]{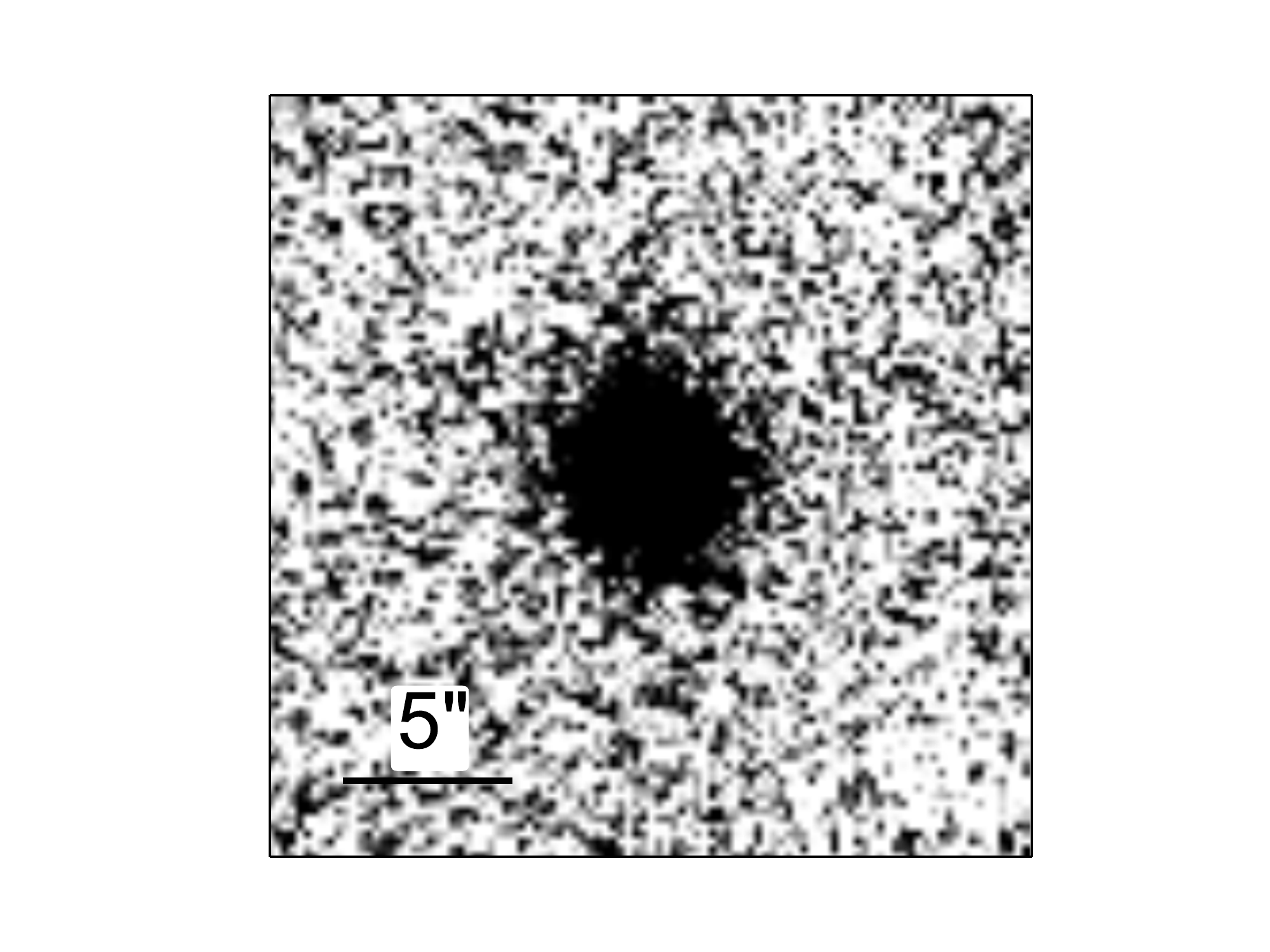}\\
 \includegraphics[clip, trim=2.5cm 1.5cm 1.5cm 1.4cm,width=0.35\textwidth]{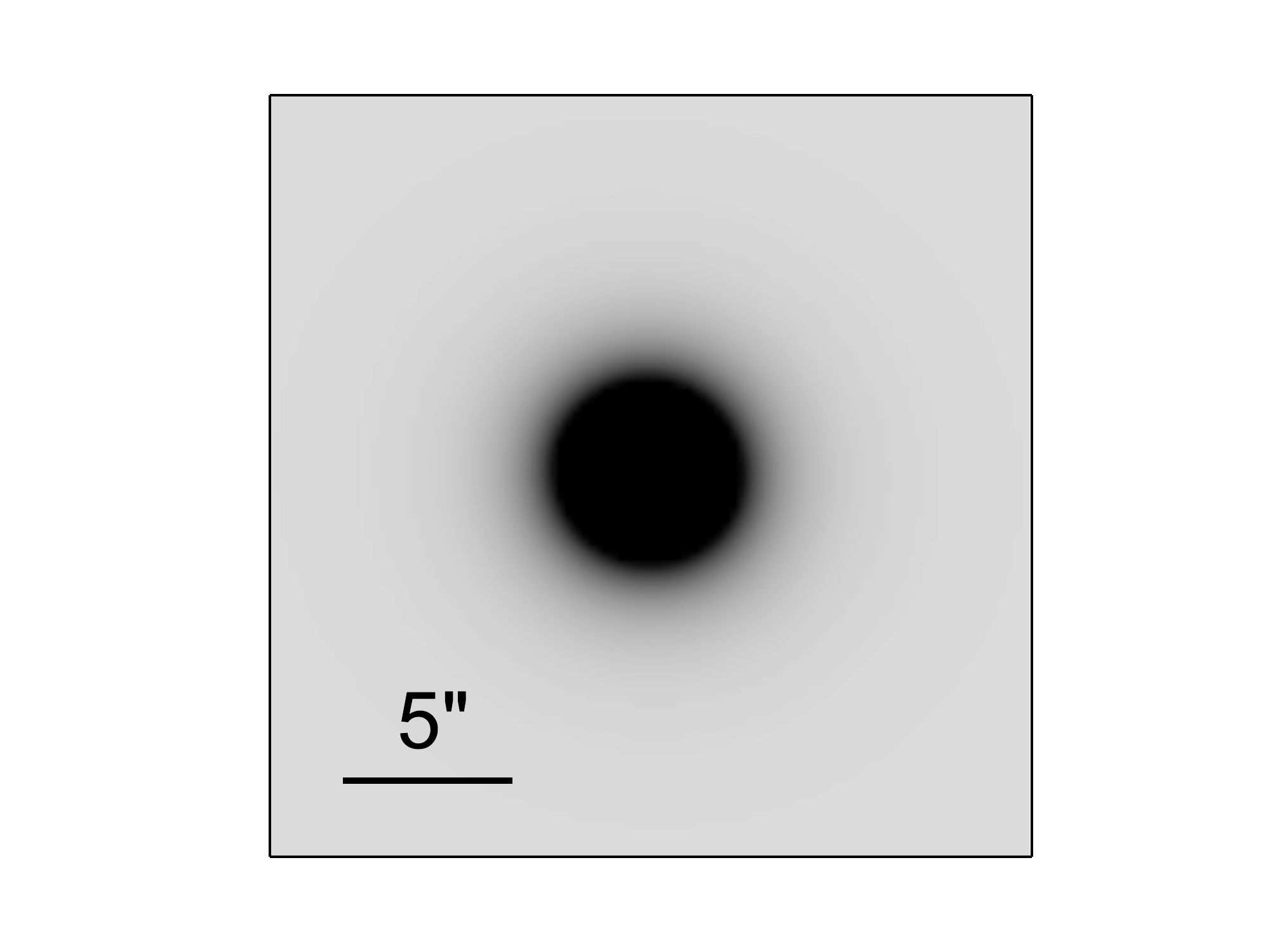} & \includegraphics[clip, trim=1.5cm 1.5cm 2.5cm 1.4cm,width=0.35\textwidth]{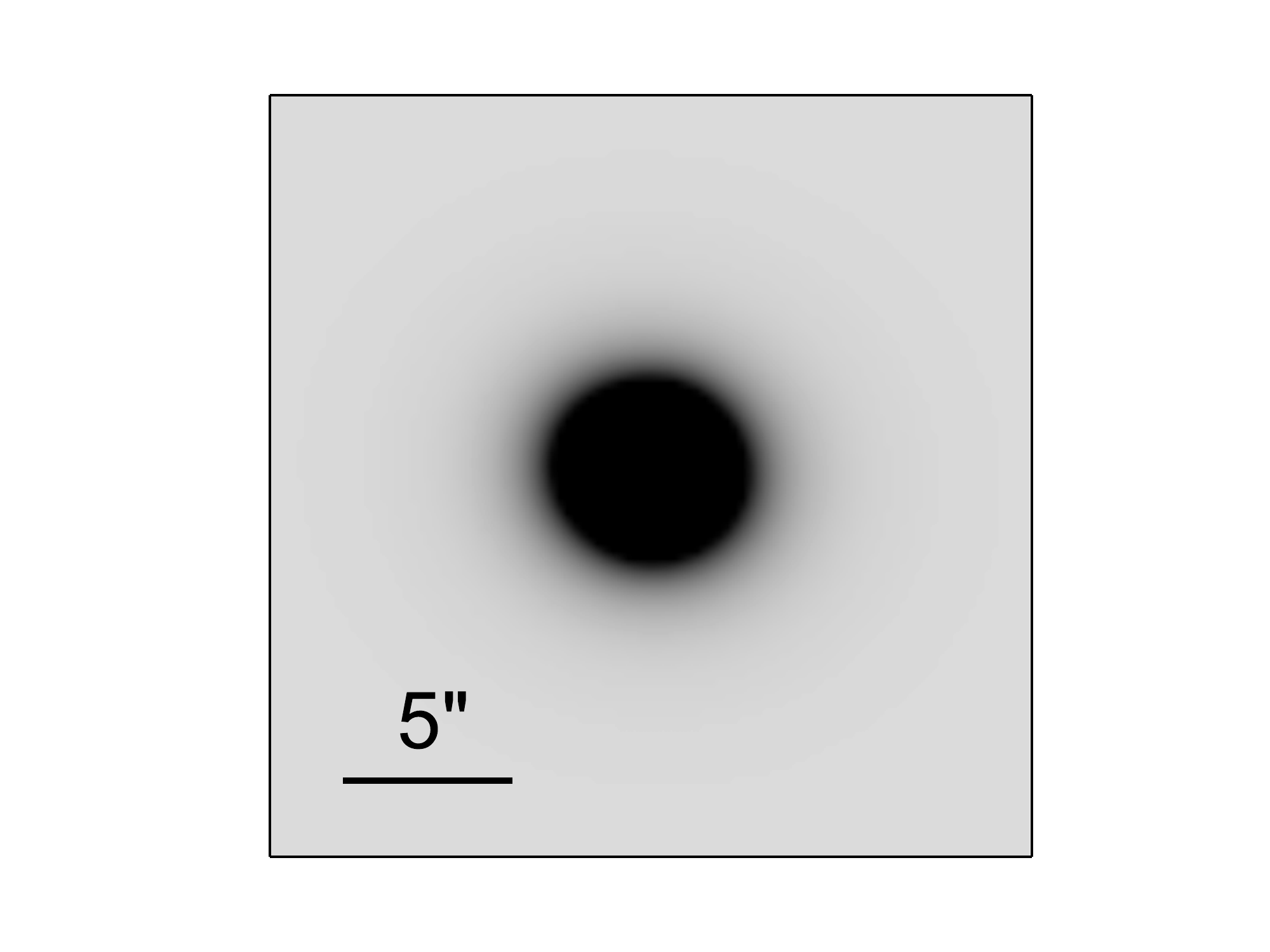}\\ 
  \includegraphics[clip, trim=0.cm 0.cm -.2cm 0.cm,width=0.4\textwidth]{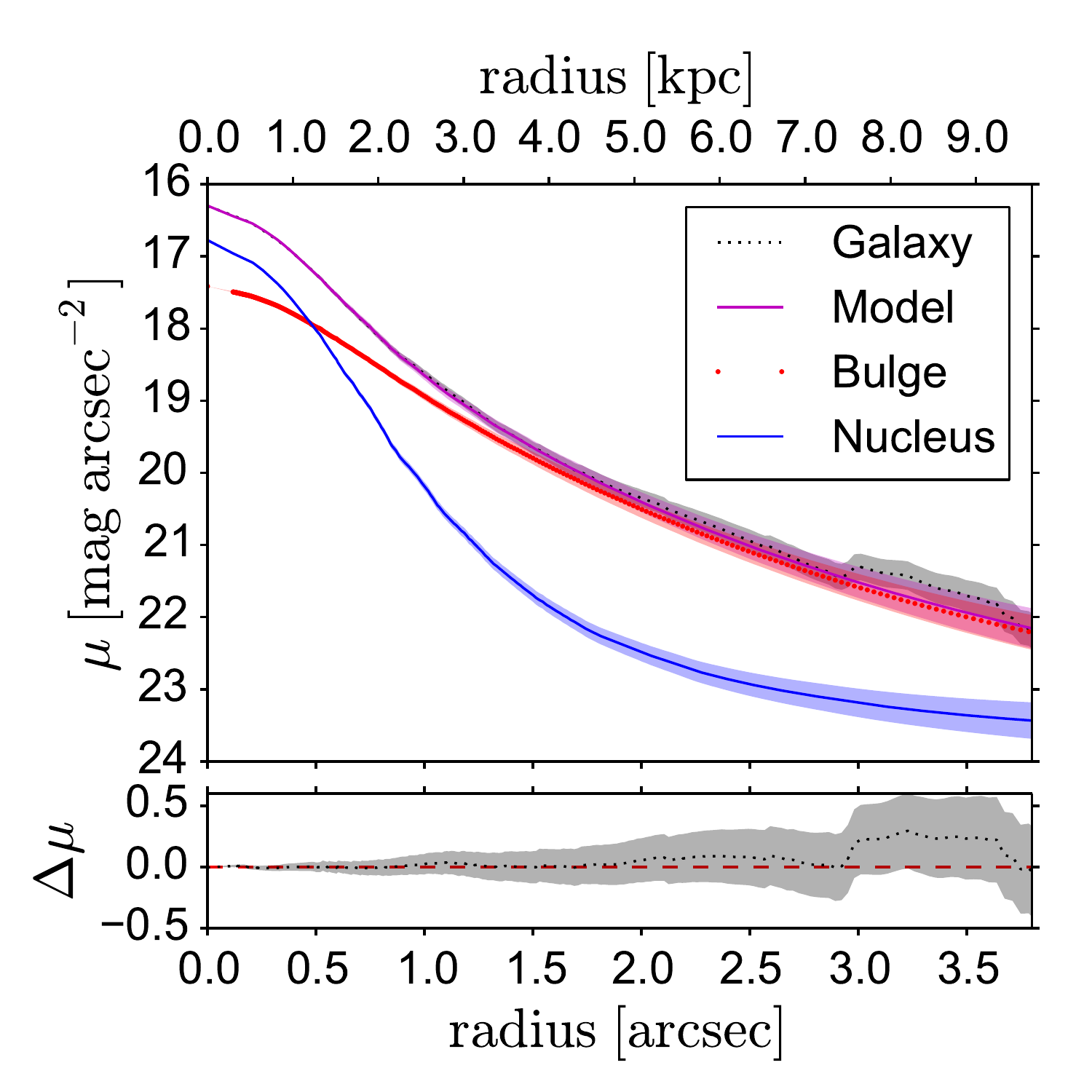} & \includegraphics[clip, trim=-.60cm 0.cm 0.4cm 0.cm,width=0.4\textwidth]{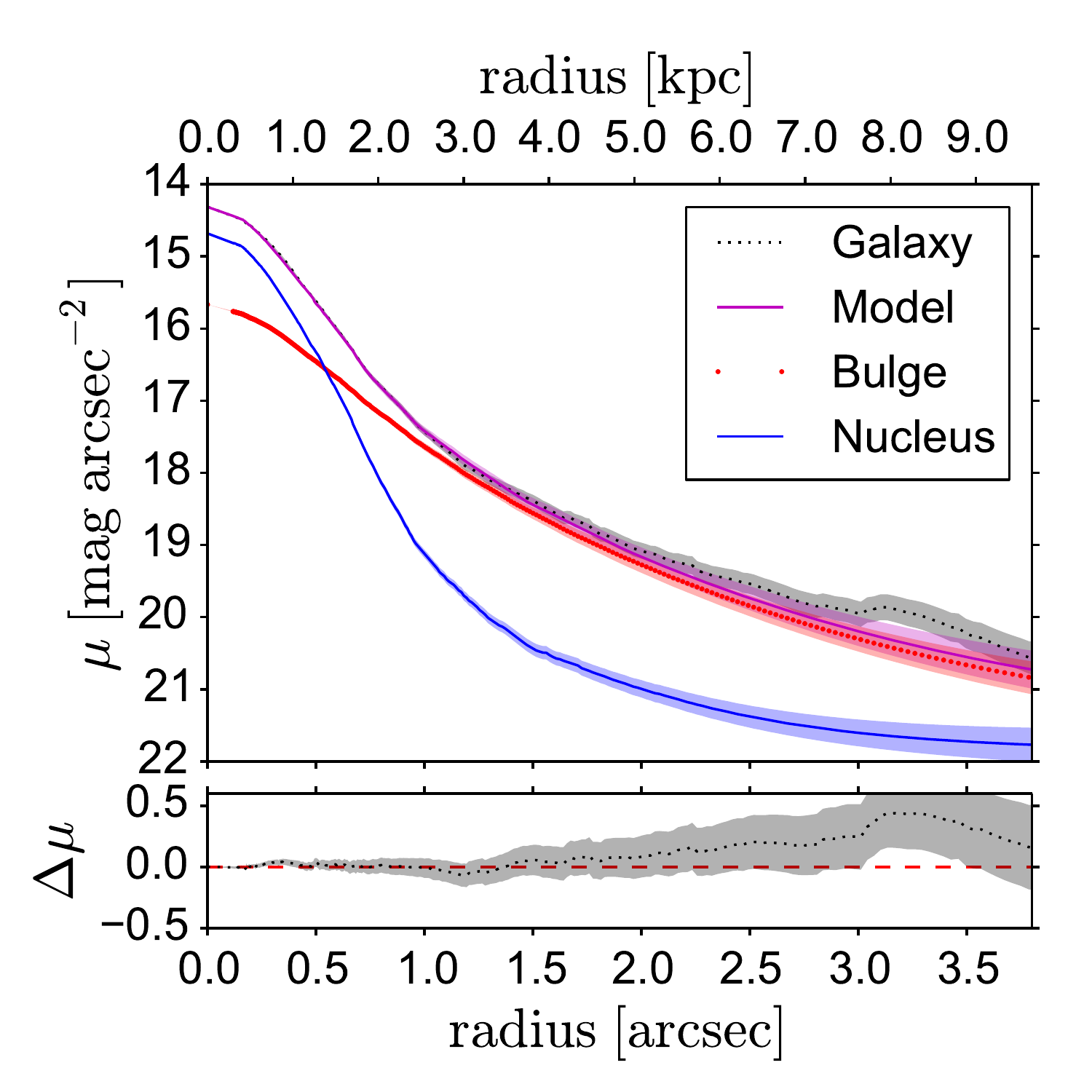}\\
  \end{tabular}
  \caption{Model 2 (AGN$+$Bulge). North is up and east is to the left. Left column shows the $J$--band and right column shows the $K_s$--band. Top row shows the observed images. Middle row shows the models images. Lower row shows the azimuthally averaged surface brightness profiles of the target, the model and the subcomponents of the model (top panel) and its residuals (bottom panel). Symbols are explained in the plots.}
   \label{sersic_model}
  \end{figure*}

Model magnitudes are also affected by uncertainties in the zero-point, estimated from magnitudes retrieved from 2MASS. Thus, zero-point magnitude variations ($\pm0.1\mathrm{mag}$) are also added as errors in the magnitudes of our final models.

\subsection{Fit of the isophotes}
Additionally to the morphological decomposition, we perform an analysis based on the ellipse fit to the galaxy isophotes \citep{wozniak_1995,knapen_2000,laine_2002,sheth_2003,elmegreen_2004,marinova_2007,barazza_2008}. We perform this analysis using the $ELLIPSE$ task in IRAF. This procedure reads a 2--dimensional image to fit isophotes to its light distribution. The fits start from an initial guess of $x$ and $y$ center, ellipticity ($\epsilon$) and position angle (PA). Each extracted isophote is represented by its surface brightness ($\mu$), semi--major axis length ($R$), PA and $\epsilon$.\smallskip

The fitted isophotes are used to represent and analyse the azimuthally averaged surface brightness profiles of the galaxy and the models derived from the photometric decomposition. Furthermore, the sample of isophotes extracted are used to identify changes in PA and ellipticity that could be associated to different structures within the galaxy morphology.\smallskip

\section{Structure of \so}
\begin{table*}
\centering
\begin{tabular}{lccccccccccc}
\hline
\hline
\multirow{2}{*}{Parameter} & \multicolumn{2}{c}{Model 1}& &\multicolumn{2}{c}{Model 2} & & \multicolumn{2}{c}{Model 3} &&  Model $4^a $&\\
\cline{2-3}
\cline{5-6}
\cline{8-9}
&	$J$	&		$K_s$	&&	$J$	&		$K_s$	& &  $J$     &$K_s$		&& $K_s$ &	   \\
\hline
\multirow{2}{*}{$m_{AGN}$} &15.65 &14.28 & &16.67 &14.39 &&16.67 &14.80&& 14.38 &\\
&(0.33) &(0.34) & &(0.38) &(0.41) &&(0.43) &(0.38)&&(0.24) &\\
\multirow{2}{*}{$m_{bulge}$} & -		&-		   && 15.70			&13.86		   &&17.77			&14.55		   && 14.97 &\\
& -        &- & & (0.37)        &(0.39) 		   &&(0.40) 			&(0.39)		   && (0.32) &\\
\multirow{2}{*}{$m_{disc}$}&- &- & &- &- &&16.28		      &14.33         && 14.90&\\
		  &- &- & &- &- &&(0.35)         &(0.38)        && (0.25) &\\
\multirow{2}{*}{$m_{bar}$} &- &- &&- &- &&-		      &-         && 15.32&\\
		  &- &- & &- &- &&-         &-        && (0.42) &\\		   
\multirow{2}{*}{$R_{eff}$ [''/kpc]} & - & - & & 0.95/2.40 & 0.82/2.07 &&0.38/0.96& 0.30/0.76&&0.43/1.10& \\
&- &- &	&(0.23/0.58) &(0.27/0.68) &&(0.13/0.32)&(0.14/0.35)&&(0.14/0.34)& \\
\multirow{2}{*}{$h_r$ [''/kpc]}&- &- & &- &- &&2.62/6.65     &3.04/7.68    &&3.19/8.10& \\
			   &- &- & &- &- &&(0.45/1.14)   &(0.40/1.01)  &&(0.47/1.20)& \\
\multirow{2}{*}{$n_{bulge}$} & -  & - && 2.58  & 2.71 &&1.80  &1.95  && 1.90 & \\
	&- &-&&(0.40) &(0.42)&&(0.31)&(0.38)&& (0.35)& \\
\multirow{2}{*}{$n_{bar}$}& -  & - & & -  & - &&-  &-  && 1.17 & \\
	&- &-&&- &-&& - &- && (0.30)& \\
\multirow{2}{*}{$\epsilon_{bar}$} & -  & - && -  & - &&-  &-  && 0.59 & \\
	&- &-&&- &-&& - &- && (0.06)& \\
\multirow{2}{*}{$\chi^2_\nu$	}& 4.250 &    3.927    & & 1.785 &    1.645    && 1.160  & 1.300  && 1.181  &			\\	
&  (0.032)&  (0.033)  & &  (0.030)&  (0.031)  && (0.027)& (0.029) && (0.022) &			\\
\hline
\end{tabular}
\caption{Best--fit parameters for model 1 (PSF only), model 2 (PSF+bulge), model 3 (PSF+bulge+disk) and model 4 (PSF+bulge+disk+bar). Parameter errors appear in parentheses$^b$.}
\begin{tablenotes}\footnotesize
\item $a$ Model 4 is only shown for $K_s-$band, since no stellar bar is detected in $J-$band.
\item $b$ Parameter errors are estimated following procedure from section \ref{errors_section}.
\end{tablenotes}
\label{results}
\end{table*}

In order to characterize the morphology of \so, we first assume that it is hosted by an elliptical galaxy, since only these type of galaxies are known to launch powerful relativistic jets able to produce $\gamma-$rays \citep{marscher_2009}. Thus, we add a S\'ersic profile to the single PSF model that represents the AGN contribution. We constrain the S\'ersic index to $n>2.0$, given the observational evidence that the light profiles of most ellipticals and classical bulges, are better described by a S\'ersic function with $n > 2$, whereas most disk-like bulges have $n < 2$ \citep{fisher_drory_2008, gadotti_2009}.\smallskip

By means of a $\chi^2$ test, we find that the improvement of this model (hereafter model 2) is equal for the $J-$ and the $Ks-$bands ($\chi^2_{model2}/\chi^2_{model1}=0.42$ for $J$--band and $\chi^2_{model2}/\chi^2_{model1}=0.42$ for $K_s$ band). The images of the galaxy and the models, as well as the azimuthally average surface brightness profiles of the galaxy, the model and the sub-components of the model for each band are shown in Figure \ref{sersic_model}.\smallskip 

\begin{figure}
\centering
\includegraphics[clip,width=0.4\textwidth]{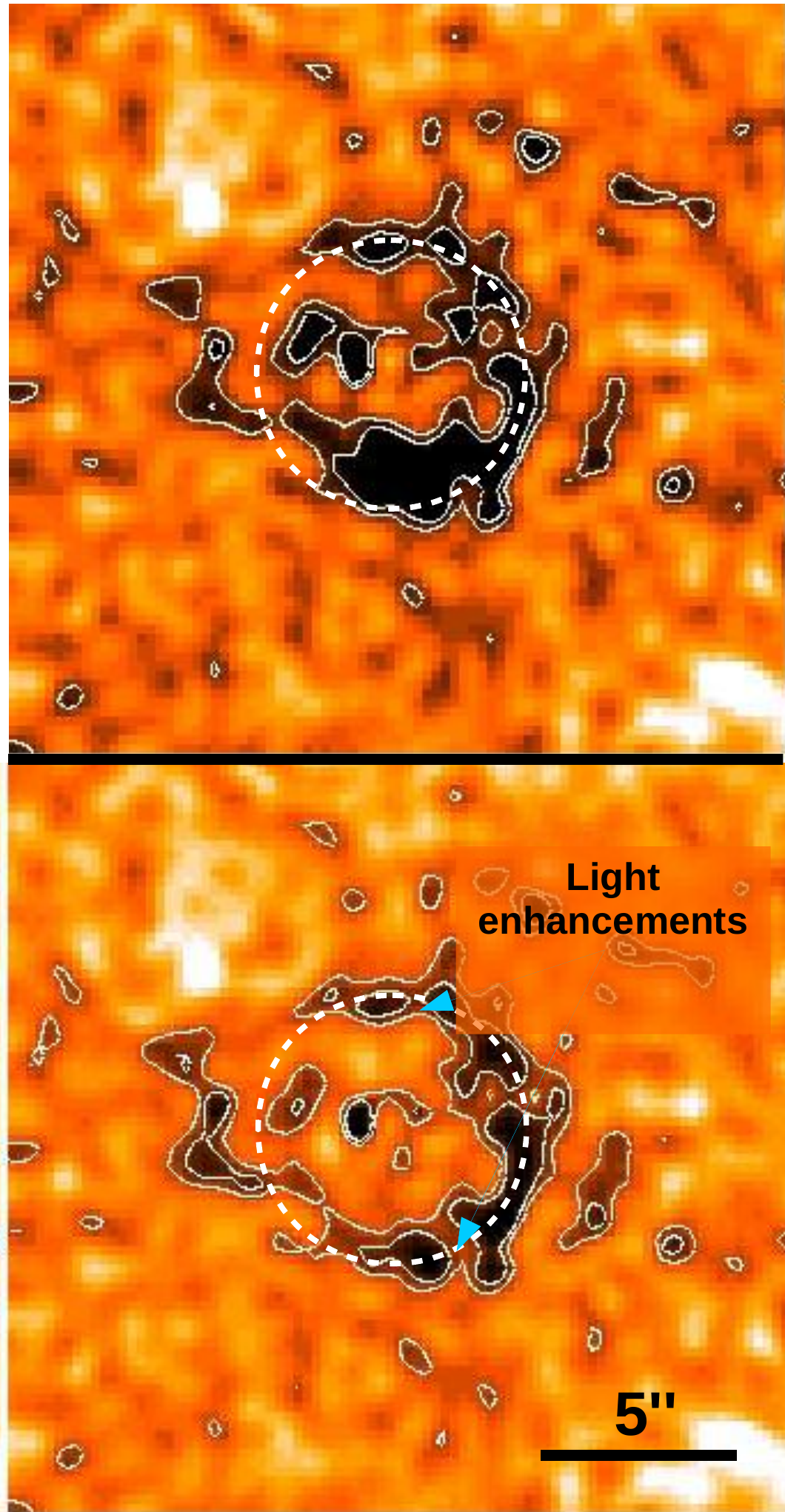}
  \caption{Residuals of the $J-$band model 2 (AGN$+$Bulge, top panel) and model 3 (AGN$+$Bulge$+$Disc, bottom panel). North is up and east is to the left. To enhance $S/N$ and to detect faint structures, the residuals were smoothed to $<1''$ resolution. The segmented white circle has a $3.2''$ radius and guides through the ring feature. Blue arrows show the light enhancements at the ends of the bar (ansae). A likely minor merger event feature is observed in the east part of the galaxy (from $R\approx3''$ up to $R \approx5''$), with a surface brightness in $J$--band $\mu=21.0\pm0.5\ \mathrm{mag}/\mathrm{arcsec}^2$, which originates the blue region at $3''$ in the $J-K_s$ colour profile of figure \ref{color}.}
   \label{residuals_j}
  \end{figure}

The residual image of the $J$--band from model 2 (top panel of Figure \ref{residuals_j}), shows a ring like feature interrupted in the eastern part. Neither the residuals nor the surface brightness profiles of the stars fitted with our PSF model show similar features. Moreover, its radius ($\sim3.5''$) exceeds by far the FWHM of our PSF ($\sim0.75''$). Hence, we consider the ring as a real component of the host galaxy.\smallskip

\begin{figure}
\centering
\includegraphics[clip,width=0.4\textwidth]{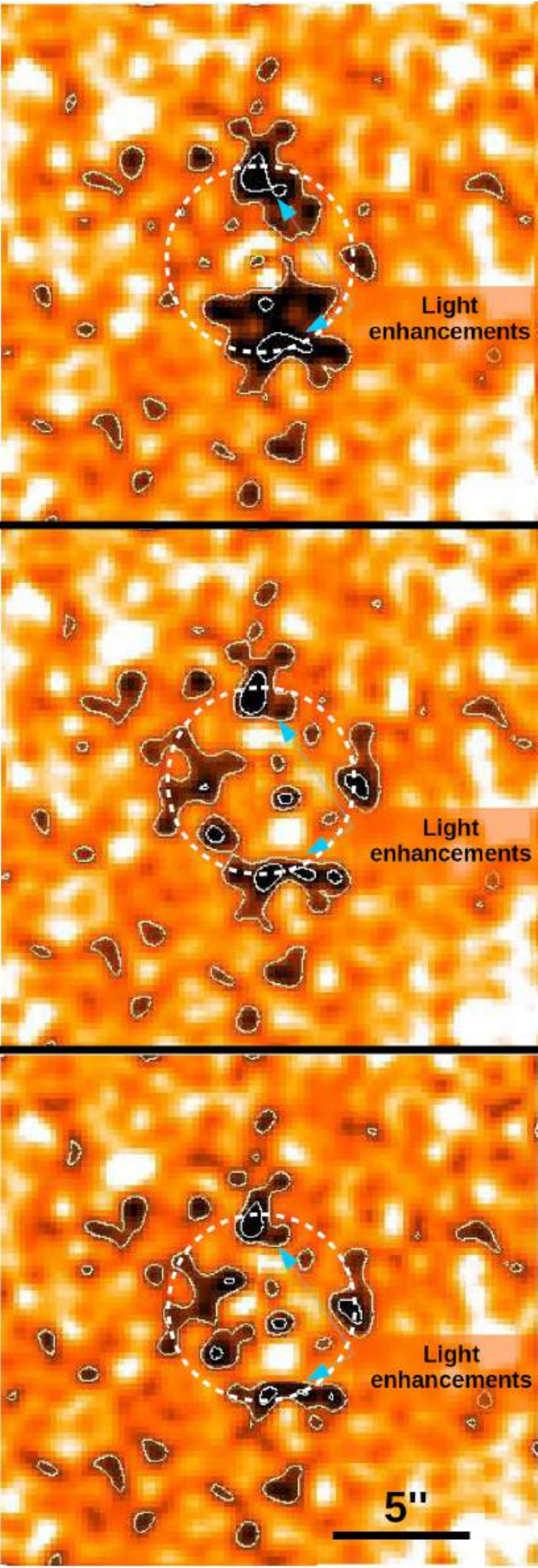}
  \caption{Residuals for the $K_s$--band model 2 (AGN$+$Bulge, top panel), model 3 (AGN$+$Bulge$+$Disc, middle panel) and model 4 (AGN$+$Bulge$+$Disc$+$Bar, bottom panel). North is up and east is to the left. To enhance $S/N$ and to detect faint structures, the residuals were smoothed to $<1''$ resolution. The segmented white circle has a $3.2''$ radius and guides through the ring feature. Blue arrows show the light enhancements at the ends of the bar (ansae). The residuals of model 2 shows hints of the bar, whereas the residuals of the models where we include a disc and a bar (model 3 and 4), show hints of the ring and the likely minor merger (eastern part of the galaxy, inside the white circle,) shown in $J-band$.  }
   \label{residuals_k}
  \end{figure}

\begin{figure*}
\centering  
\begin{tabular}{rr}
\includegraphics[clip, trim=2.5cm 1.5cm 1.5cm .4cm,width=0.35\textwidth]{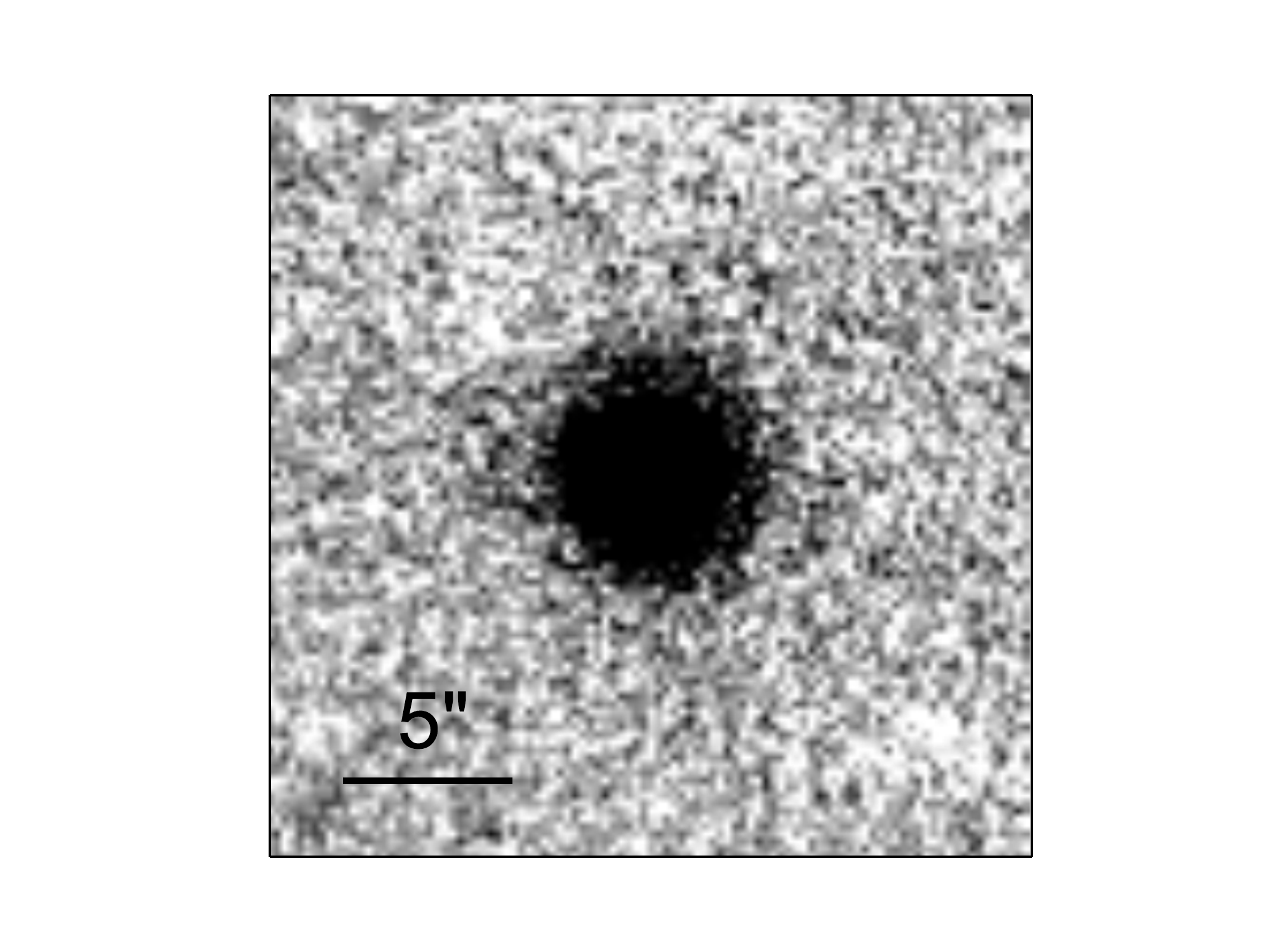} & 
\includegraphics[clip, trim=1.5cm 1.5cm 2.5cm 1.4cm,width=0.35\textwidth]{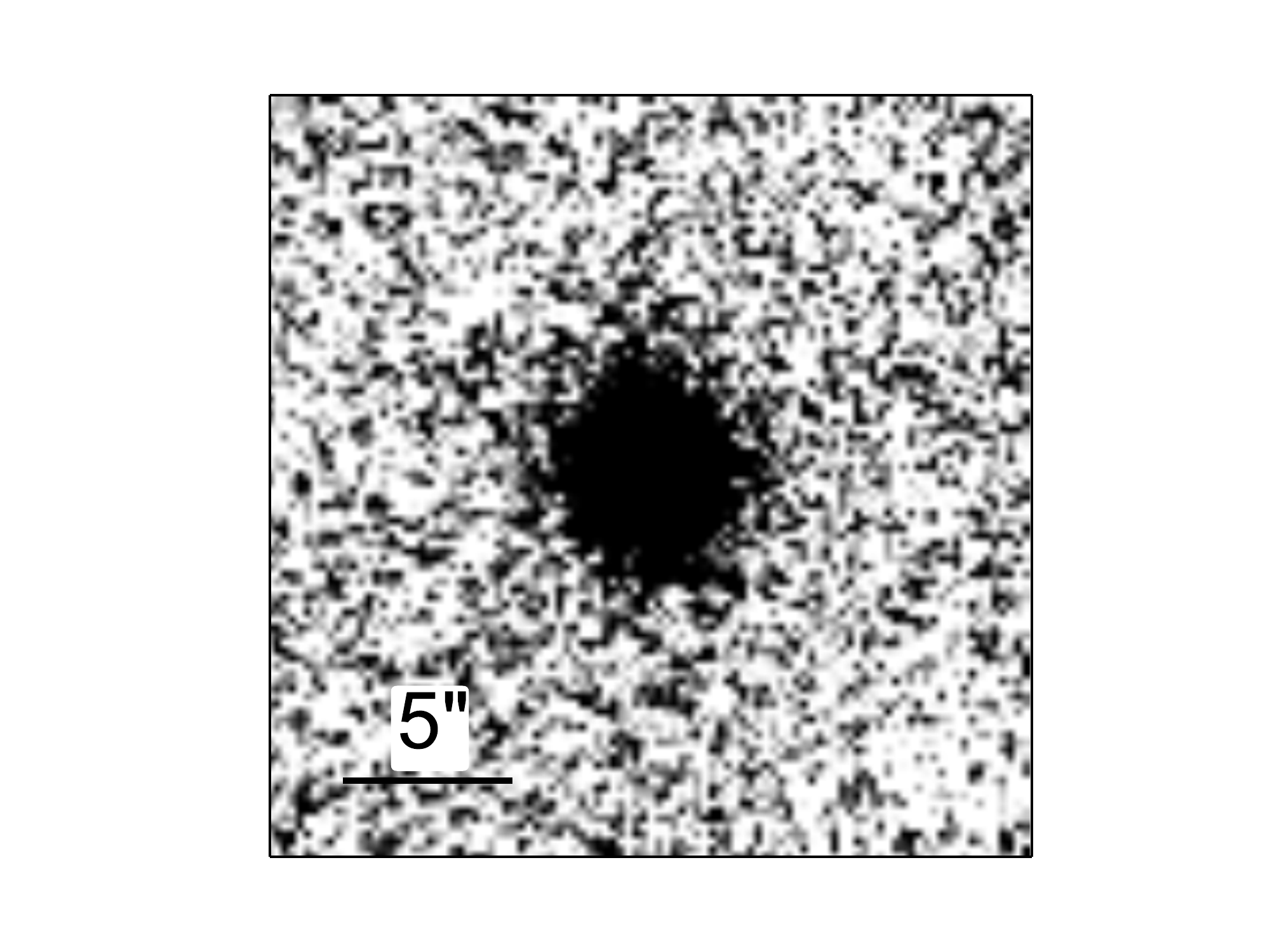}\\
\includegraphics[clip, trim=2.5cm 1.5cm 1.5cm 1.4cm,width=0.35\textwidth]{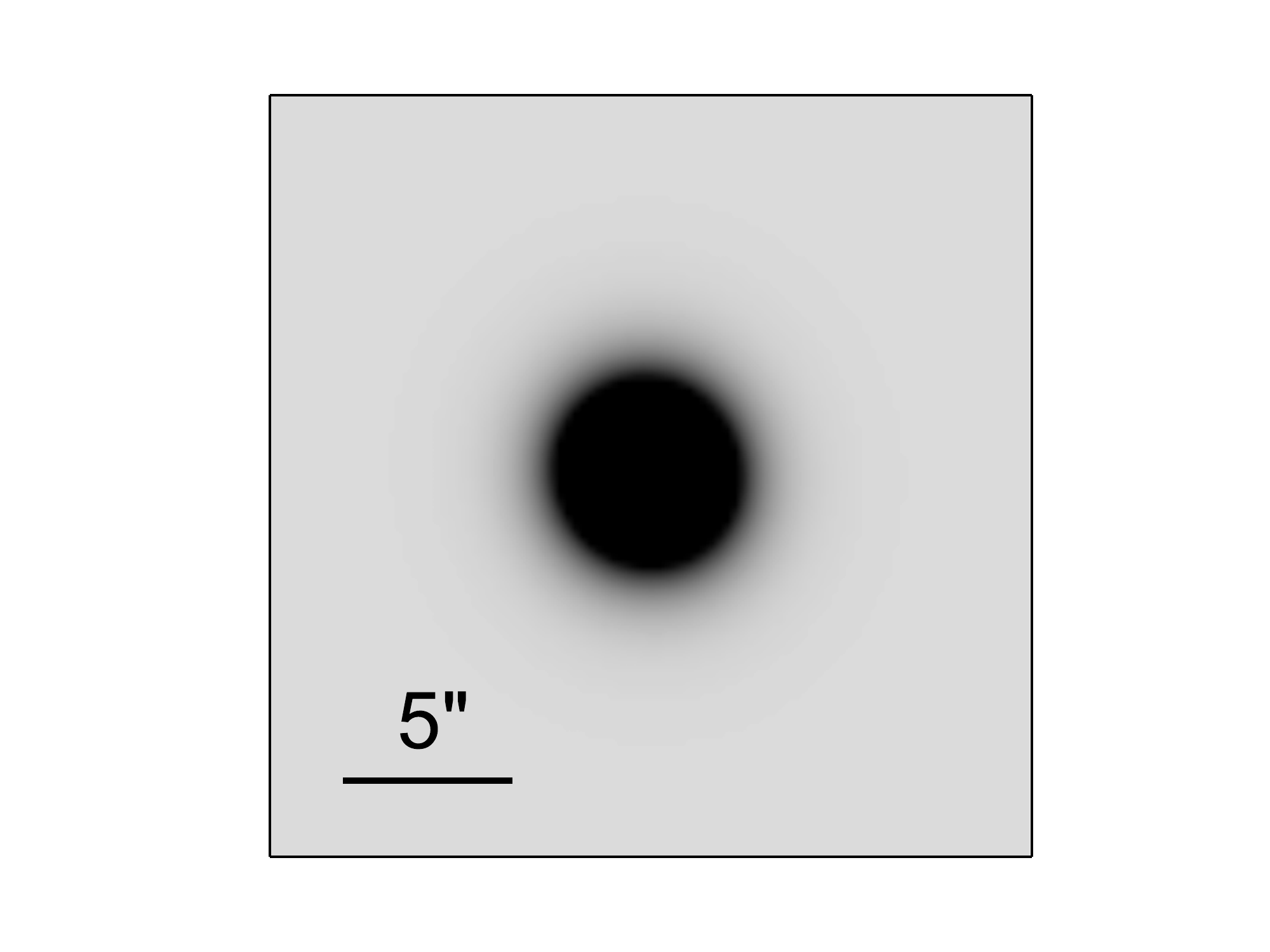} & 
\includegraphics[clip, trim=1.5cm 1.5cm 2.5cm 1.4cm,width=0.35\textwidth]{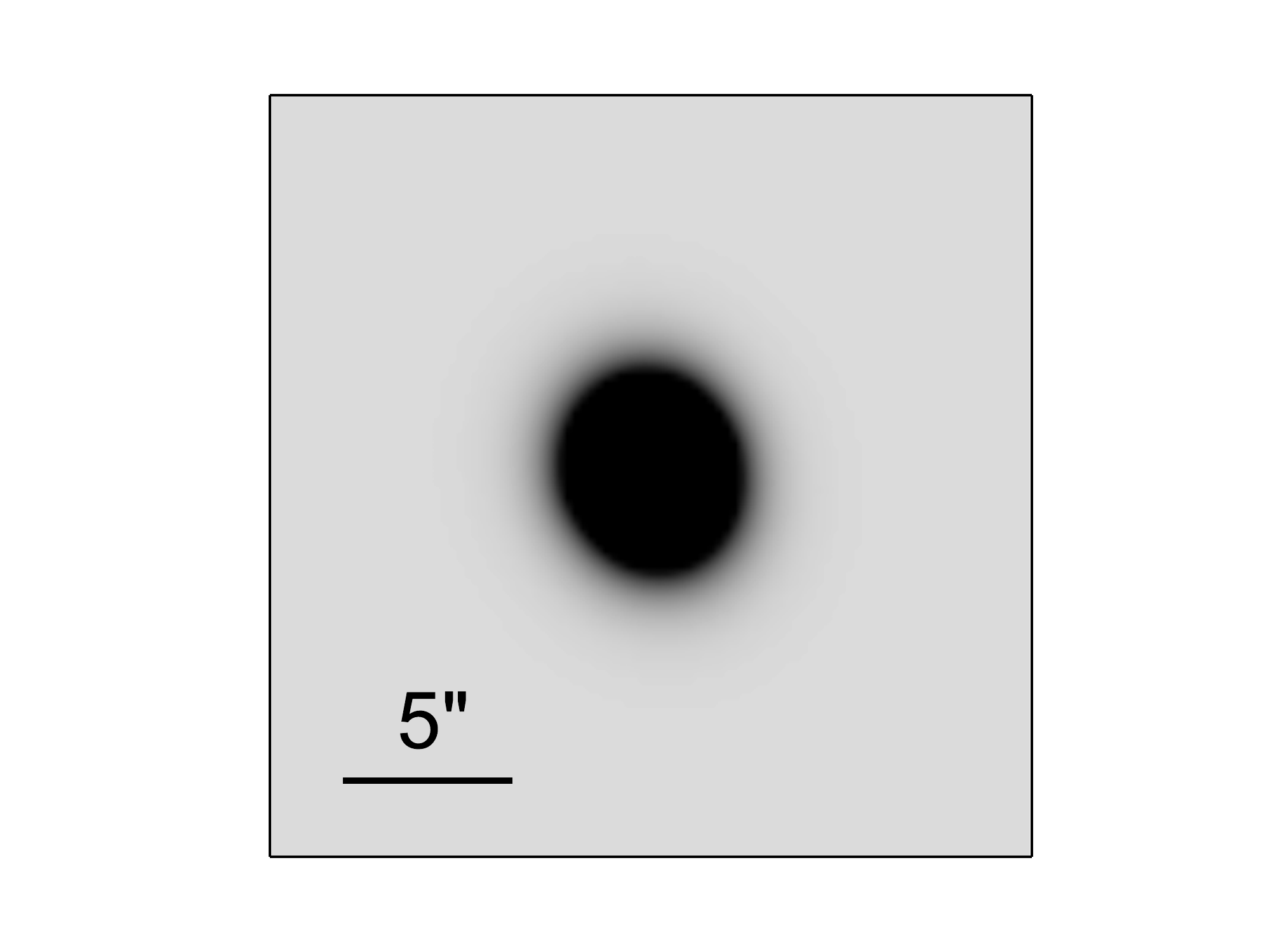}\\ 
\includegraphics[clip, trim=0.cm 0.cm -0.2cm .0cm,width=0.4\textwidth]{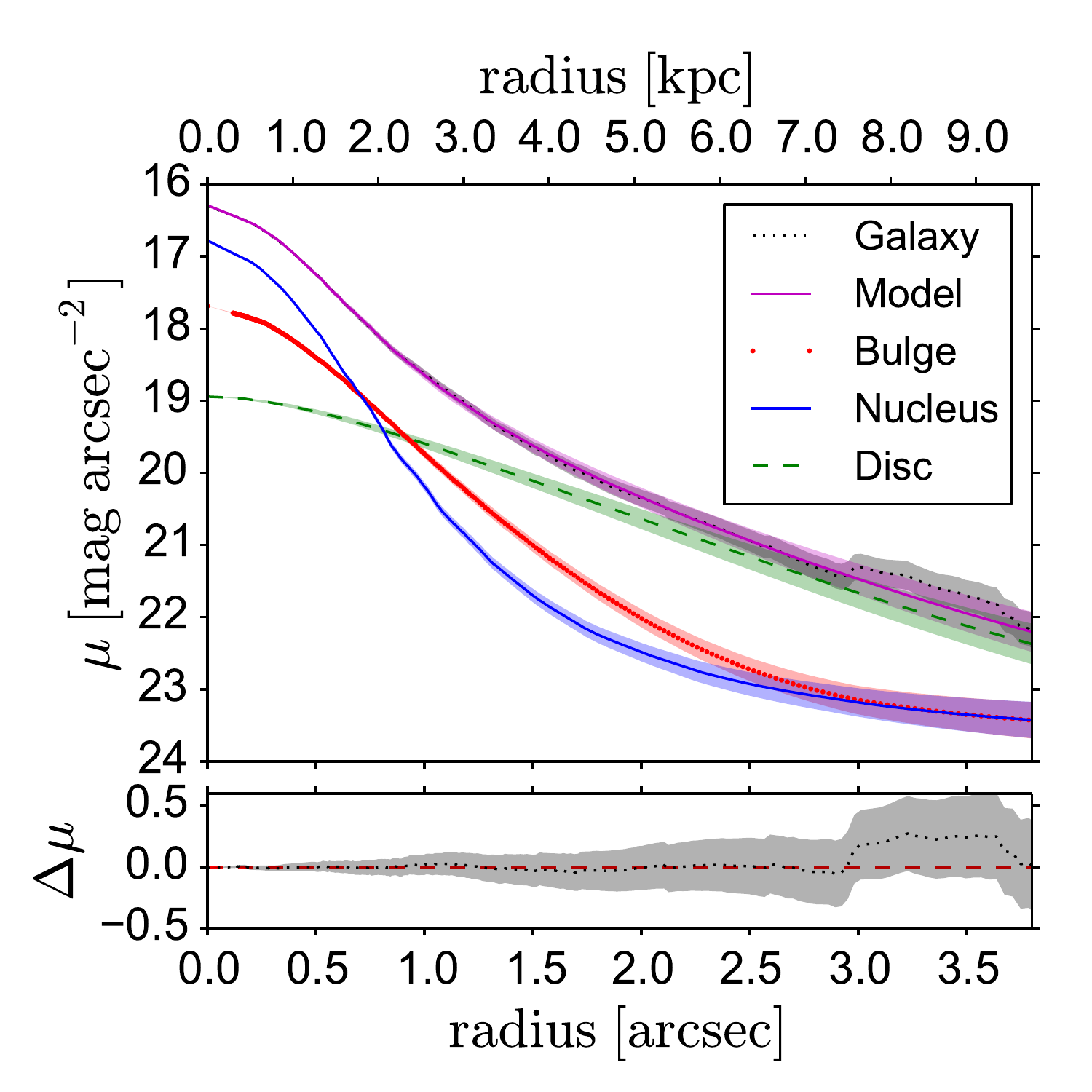} & 
\includegraphics[clip, trim=-0.60cm 0.cm 0.4cm 0.cm,width=0.4\textwidth]{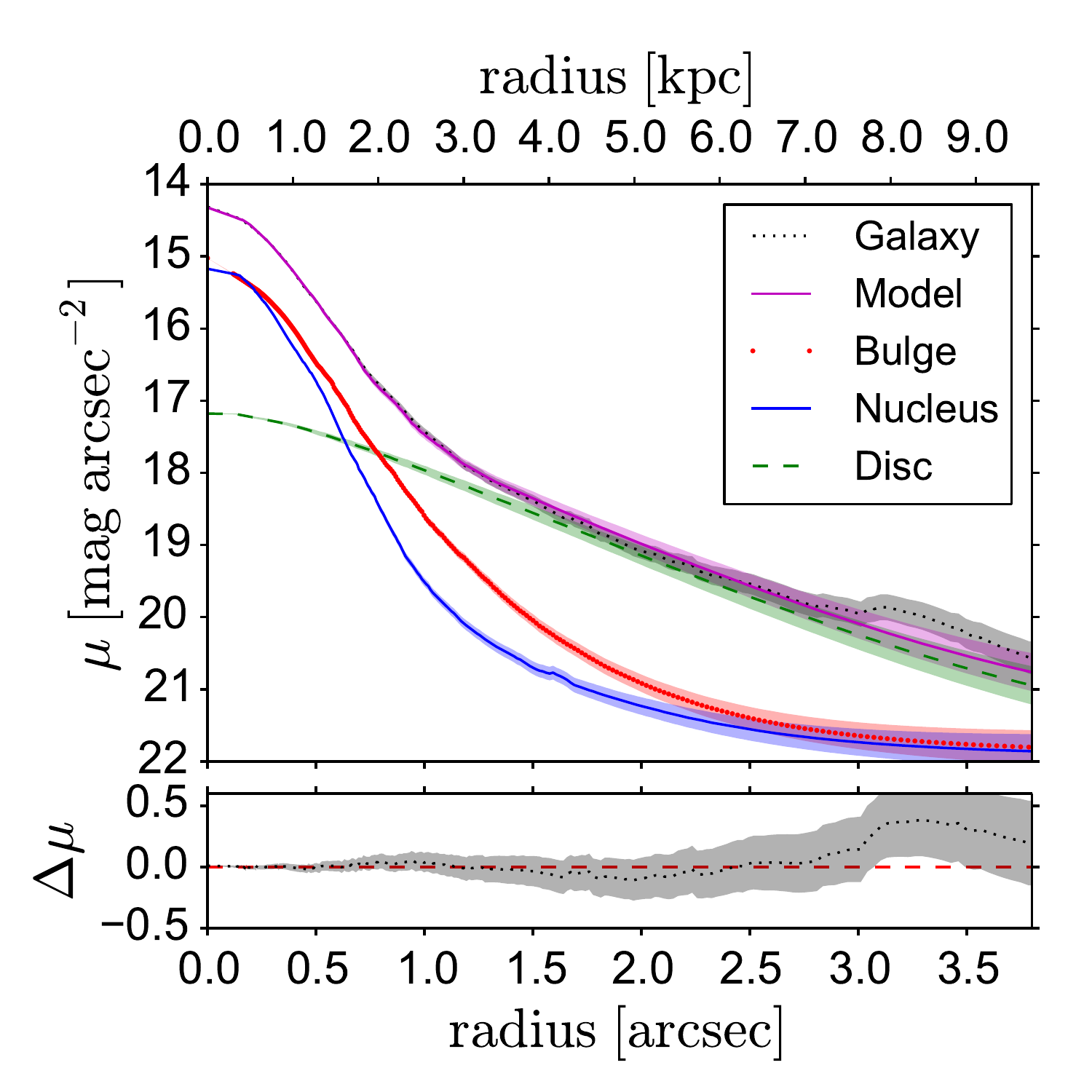}\\
  \end{tabular}
  \caption{Model 3 (AGN$+$Bulge$+$Disc) for \so. Left column shows the $J$--band and right column shows the $K_s$--band. Top row shows the observed images. Middle row shows the images of our models. Lower row shows the azimuthally averaged surface brightness profiles of the target, the model and the subcomponents of the model (top panel) and its residuals (bottom panel). Symbols are explained in the plots.}
   \label{disc_model}
  \end{figure*}

The $K_s$--band residual (top panel of Figure \ref{residuals_k}) shows an elongated and roughly symmetric structure with a length similar to the diameter of the ringed feature ($\sim3.2''/\sim8.1\mathrm{kpc}$). In both bands a not-fitted bump in the light distribution of the galaxy is observed (from $\sim2.8'' $ to $ \sim3.7'' $), which is consistent with the ring and the two light enhancements close to the ends of the elongated structure. Since residuals are still considerable, we include an extra component into the last model (Figure \ref{disc_model}). We choose an exponential function, since it is able to represent the likely presence of a disk in the host galaxy of a typical NLSy1 galaxy (we call it model 3). The improvement over model 2 is $\chi^2_{model3}/\chi^2_{model2}=0.65$ for $J$--band and $\chi^2_{model3}/\chi^2_{model2}=0.79$ for $K_s$--band. From the residual images, we observe that the ring in $J$--band (bottom panel of Figure \ref{residuals_j}) seems better defined. Moreover, in $K_s$--band (middle panel of Figure \ref{residuals_k}), hints of this structure emerge, whereas the elongated structure disappear. 

\begin{figure}
\centering  
\begin{tabular}{c}
\includegraphics[clip, trim=.1cm 5.5cm .1cm 2.5cm,width=0.40\textwidth]{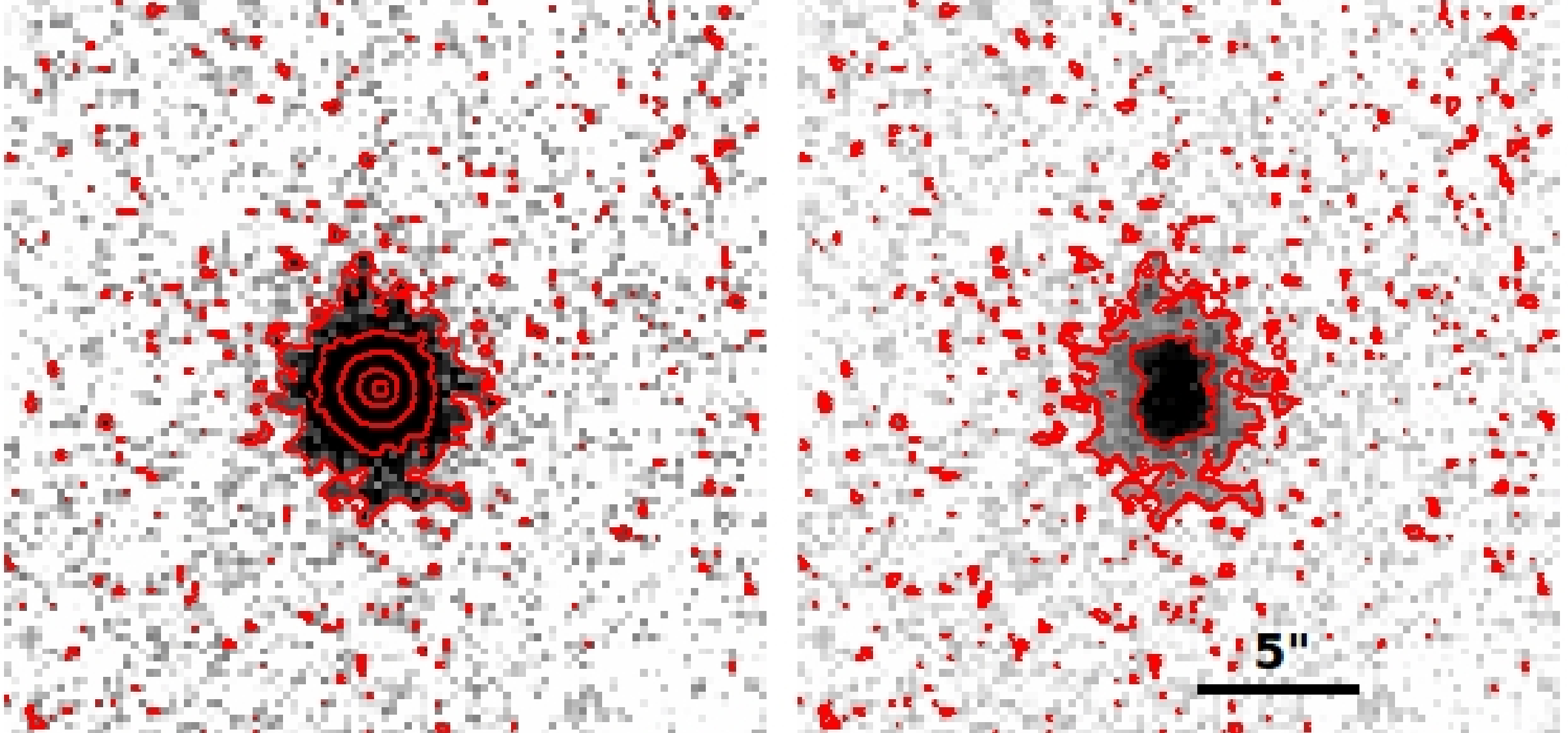} \\
 \includegraphics[clip, trim=0.85cm 0.5cm 0.5cm 1.3cm,width=0.47\textwidth]{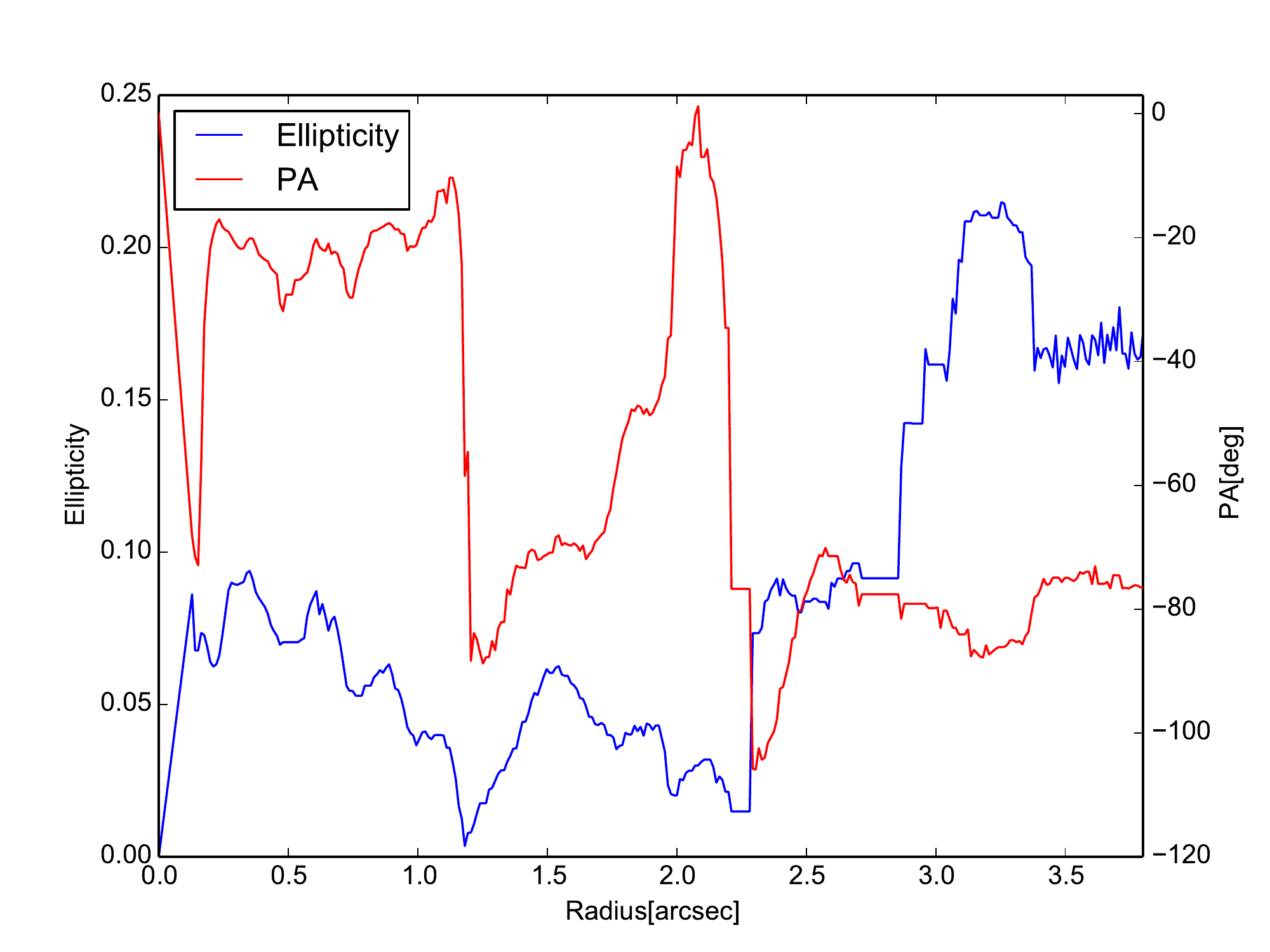} \\
  \end{tabular}
  \caption{In top panels we compare the observed $K_s$--band image (top left), with an image in $K_s$--band where the AGN and bulge models have been subtracted (top right). The bar becomes visible on the underlying disk if the AGN and bulge contributions are subtracted. In the lower panel we show the average profiles of ellipticity $\epsilon$ (blue line) and position angle PA (red line) obtained with $ELLIPSE$ from $K_s$--band plotted against the isophote major axis.}
   \label{disc_bar}
  \end{figure}

The elongated feature and the light enhancements might be explained by the presence of a stellar bar showing their ansae \citep[bright regions at the ends of bars observed in $\sim40\%$ of SB0 galaxies, as found by][]{martinez_2007, laurikainen_2007}. Such a bar could be more likely detected in $K_s-$band since neither young luminous stars or dust strongly affect its observed emission \citep{rix_rieke_1993}. Nevertheless, a powerful AGN, a bright bulge and a disc, might outshine the bar, making its presence less evident. The upper panels of Figure \ref{disc_bar}  shows an image of \so\ in Ks--band, with the AGN and bulge contribution subtracted (using a bulge+AGN+disc model), revealing an elongated and symmetrical feature that resembles a stellar bar over the underlying disk.\smallskip

\begin{figure}
\centering  
\begin{tabular}{r}
\includegraphics[clip, trim=2.5cm 1.cm 2.5cm 1.4cm,width=0.35\textwidth]{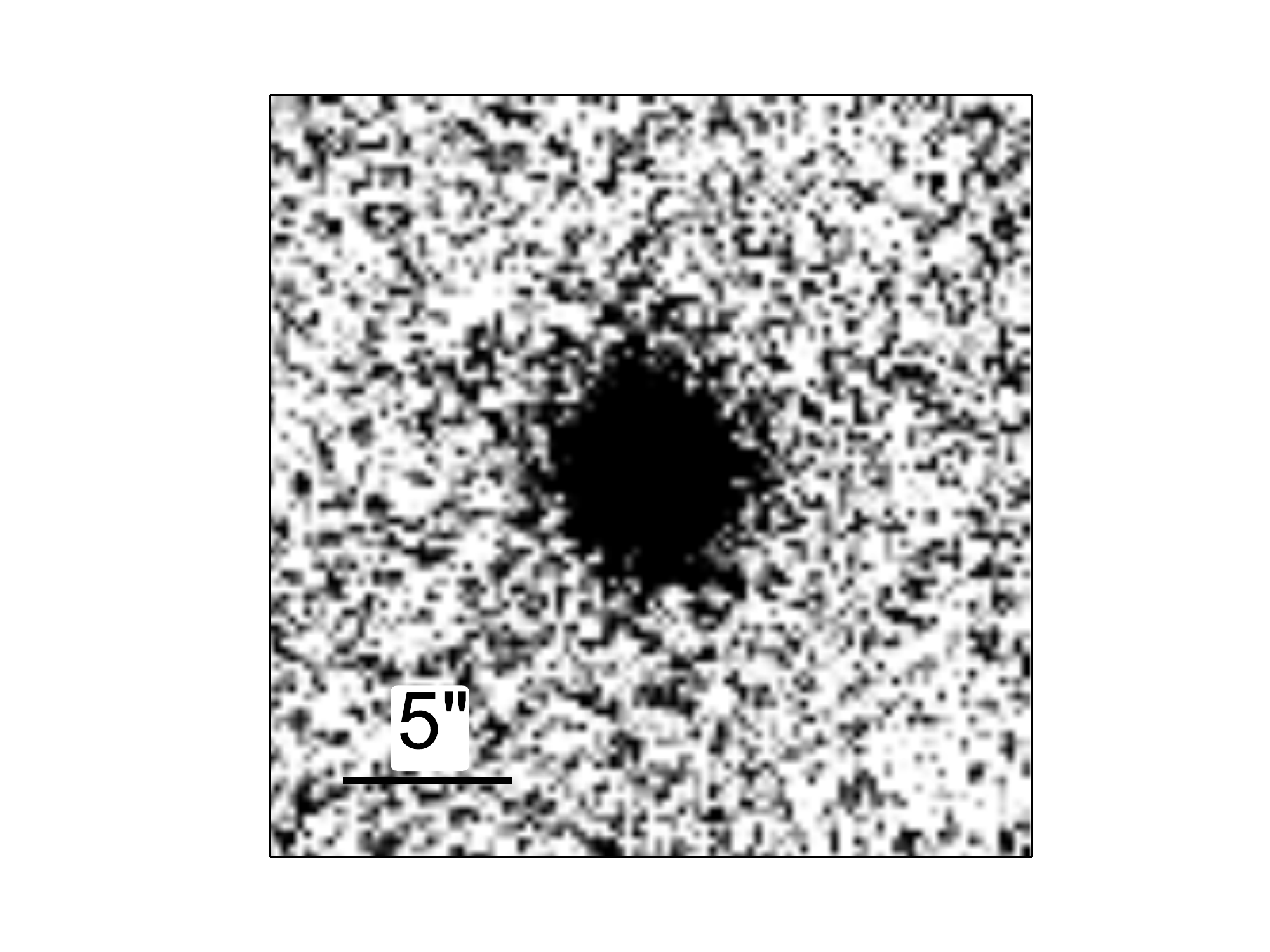} \\
 \includegraphics[clip, trim=2.5cm 1.5cm 2.5cm 1.5cm,width=0.35\textwidth]{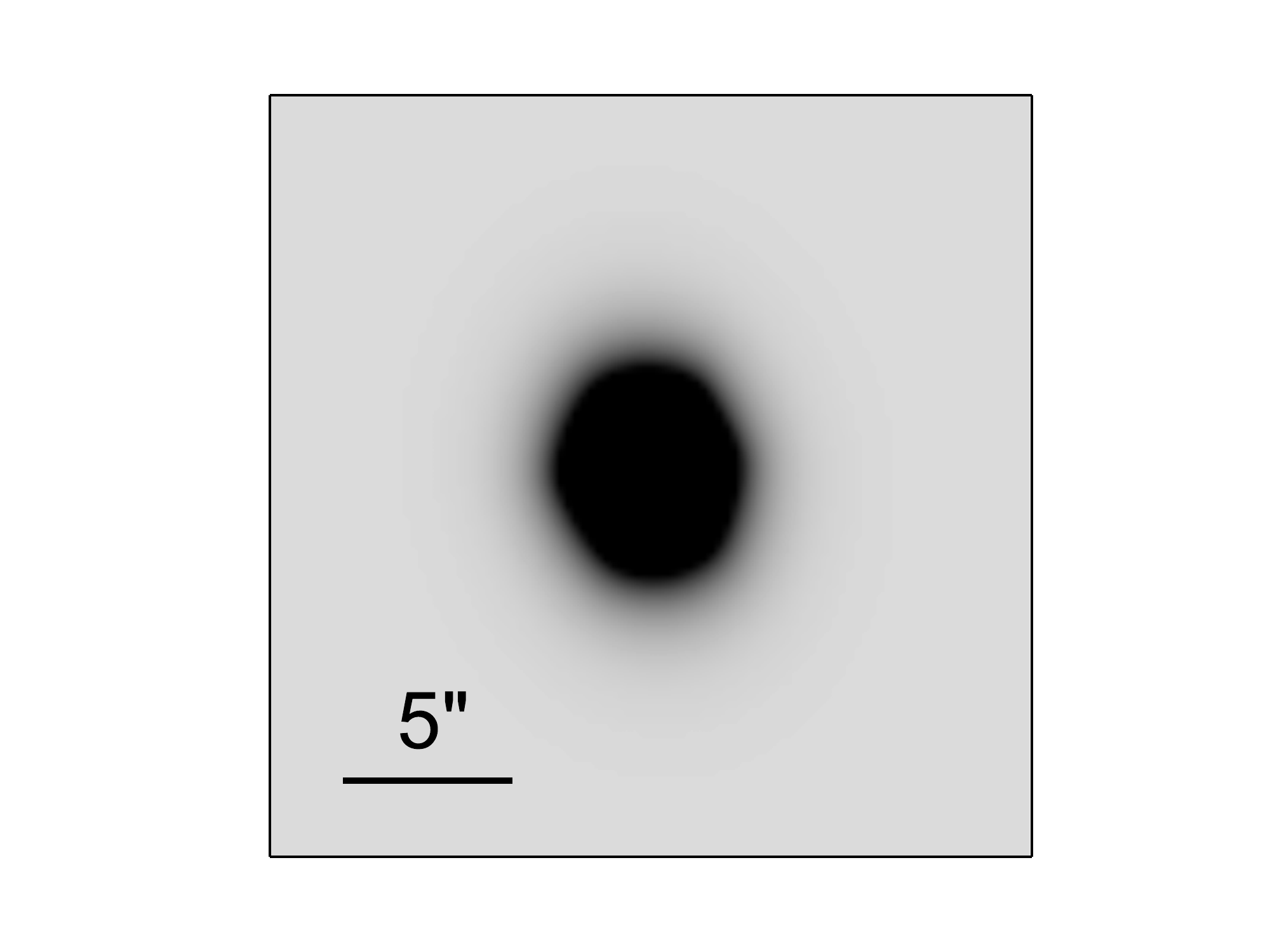} \\ 
  \includegraphics[clip, trim=0.5cm 0.6cm 0.6cm 0.8cm,width=0.4\textwidth]{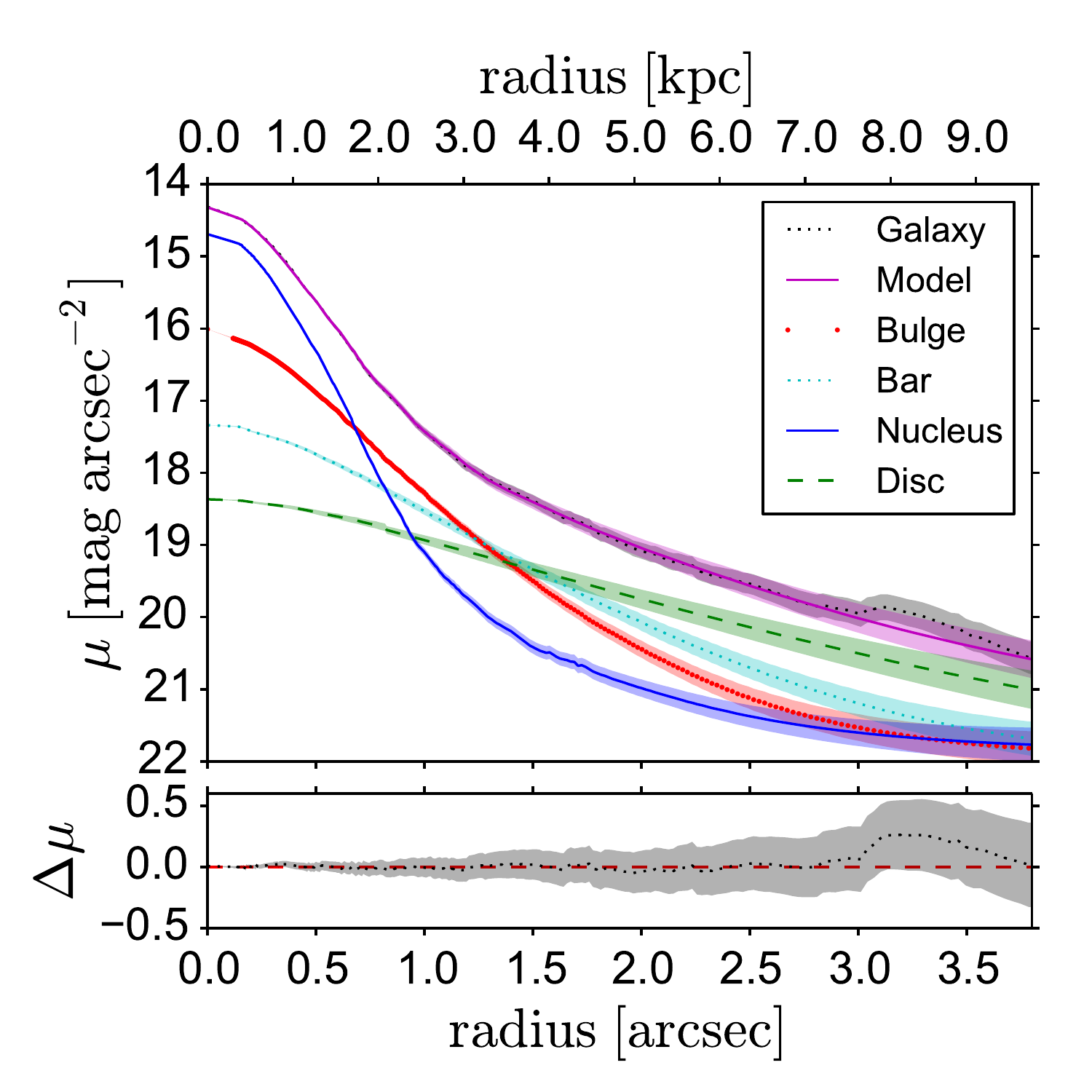} \\
  \end{tabular}
  \caption{Model 4 (AGN$+$Bulge$+$Disc$+$Bar) for the $K_s-$band. Top panel shows the observed image. Middle panel shows our model image. Lower row shows the azimuthally averaged surface brightness profiles of the target, the model and the subcomponents of the model (top panel) and its residuals (bottom panel). Symbols are explained in the plots.}
   \label{bar_model}
  \end{figure}


In order to confirm the existence of a bar in the host galaxy of \so, we perform another widely used method for detecting and describing bars; the ellipse fit of the galaxy isophotes (see plot in the lower panel of Figure \ref{disc_bar}). When $\epsilon$ and PA are plotted against radius, a bar is characterized by a local maximum in $\epsilon$ and a constant PA (typically $\Delta \textrm{PA}\lesssim 20^\circ$) along the bar \citep{wozniak_1995,jogee_1999, mendendez_2007}. We can see a region that fulfil these criteria (from  $ \sim2.6'' < \textrm{radius} <\ \sim3.2''$ and PA $\sim78^\circ$) suggesting again the presence of a bar. Since the method of the ellipses fit also hints at the presence of a bar, we proceed to characterized its morphology (Figure \ref{bar_model}). We add a S\'ersic profile to model 3 of $K_s$--band to fit the light distribution of the stellar bar (we call it model 4). We use as initial guesses a S\'ersic index $n=0.5$ \citep{bar_2008} and the $\epsilon$ and PA derived from the ellipse fit of Figure \ref{disc_bar}. The improvement with respect to the model where no bar is included is $\chi^2_{model4}/\chi^2_{model3}=0.90$. From the residual image (shown in lower panel of figure \ref{residuals_k}) we can see that in general, the residuals decrease, the hints of the ring remain and the ansae are better defined. The bump remains unfitted with the functions included in the model, which is expected given that it is caused by a ring and the bar ansae. The parameters derived from every model analysed are shown in table \ref{results}.\smallskip




\subsection{NIR colour gradient}\label{sec:NIR_color}
\begin{figure}
\includegraphics[width=0.5\textwidth]{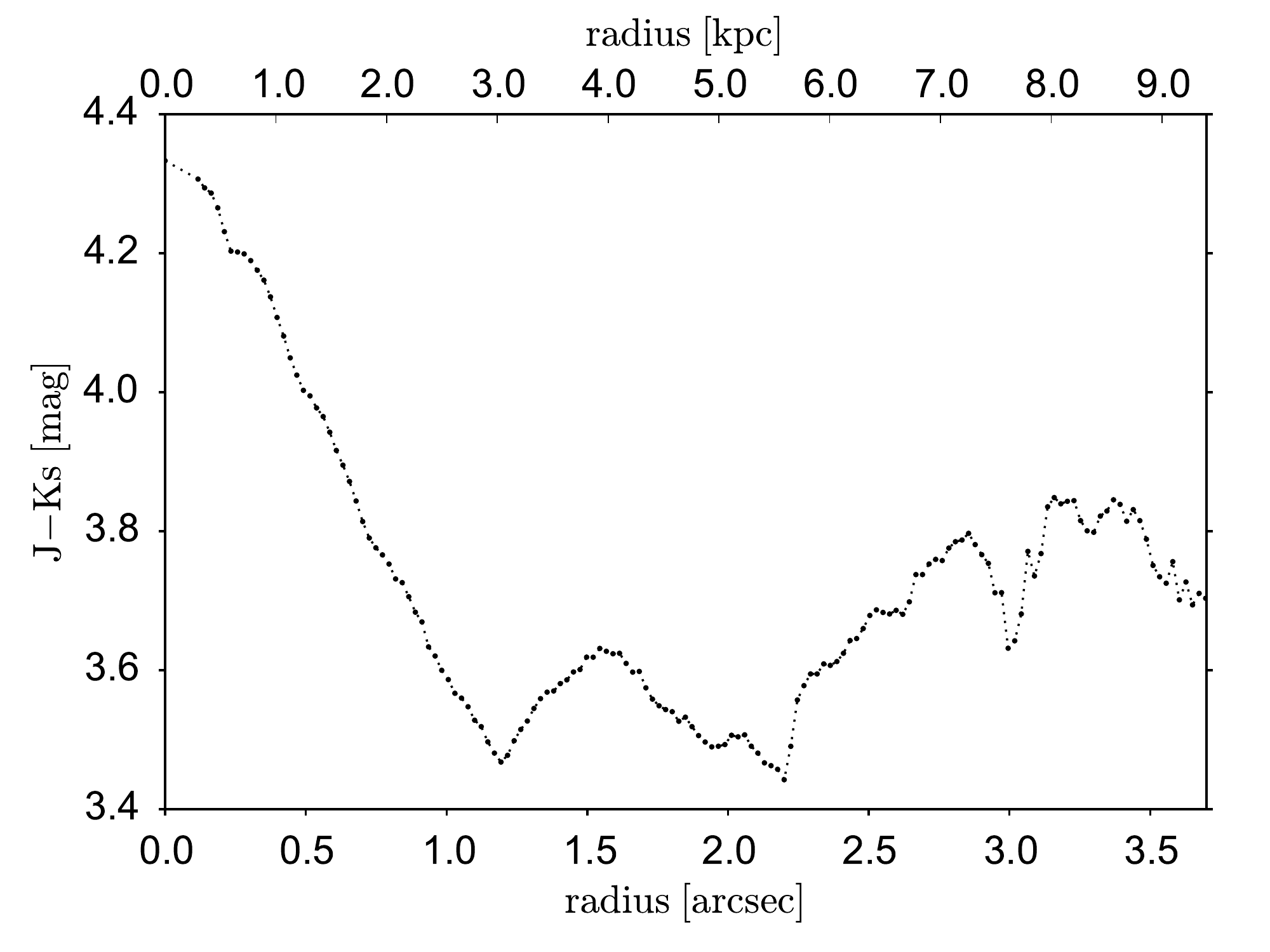}
\caption{ 
The radial $J-Ks$ colour profile of the host galaxy of \so. The AGN contribution has been subtracted from the best fit model of each band.
}

   \label{color}
\end{figure}

Figure \ref{color} shows the $J-K_s$ colour profile of the host galaxy of \so. The AGN contribution has been subtracted using the best fit model for each band.\smallskip
In general, as we move from the center to the outer parts of the host galaxy, we can see that the colour decreases from $ J-K_s=4.33\ \mathrm{mag} $ down to $J-K_s=3.45\ \mathrm{mag}$ at $R=1.20''$, showing that the central region (bulge) is the reddest of the host galaxy. From $R=1.20''$ towards larger radii, the colour increases up to a local maximum of $J-K_s=3.63\ \mathrm{mag}$ at $R=1.55''$. We link this increase in color to the bar, since here is where it has its maximum influence (see Figure \ref{bar_model}). A second increase in colour is observed in the bar region from $R=2.20''$ to $R=2.85''$, with a maximum of $J-K_s=3.80\ \mathrm{mag}$. Between $2.85'' < R < 3.15'' $, a blue region is observed which corresponds to eastern feature (inside the ring) in figures \ref{residuals_j} and \ref{residuals_k}. We observe a last local minimum at $R=3.30'' $ with a colour $J-K_s=3.80\ \mathrm{mag}$. We associate this colour to the ring, with no influence of the ansae since, according to observations by \citep{martinez_2007}, ansae do not show any colour enhancement (probably because they are a dynamical phenomena). Finally, as we move outward, the disk becomes bluer, reaching an average colour $J-K_s=3.70\ \mathrm{mag}$.\smallskip

\section{The host galaxy of \so}\label{section:host_galaxy}

According to the results shown in table \ref{results}, the host of \so\ can be classified as a barred lenticular galaxy (SB0). In addition to the ansae morphology, that is frequent in S0 galaxies ($\sim40\%$ of S0) as found by \citet{laurikainen_2007}, both the bulge and the disc fulfil the characteristics for lenticular galaxies presented in  \citet{laurikainen_2010}. They also find that, as in spirals \citep{hunt_2004,noordermeer_2007}, the luminosity of the bulge in S0s correlate to the luminosity of their discs. According to such correlation ($M_{K,disk}=0.63M_{K,bulge}+9.3$), the bulge of \so\ should have a disk with an absolute magnitude $ M_{K,disk}=-24.55\pm0.20 $, consistent with the absolute magnitude derived through the morphological analysis in this work ($ M_{K,disk}=-24.85\pm0.25 $).\smallskip 

The parameters derived in this work for the components of \so\ are consistent with those of pseudobulges. \citet{weinzirl_2009} find a connection with pseudobulges and small S\'ersic indeces ($n<2.0$), consistent with $n=1.8\pm0.31$ for $J$--band and $n=1.9\pm0.35$ for $K_s$--band, derived for \so. Independently, \citet{fisher_drory_2008} find that pseudobulges and their discs are associated through their effective radius and scale lengths as $r_{eff}/h_r=0.21\pm0.10$ consistent with \so\ ($r_{eff}/h_r=0.14\pm0.07$ for $J$--band and $r_{eff}/h_r=0.14\pm0.06$ for $K_s$--band). On the contrary, they found that classical bulges have large $r_{eff}/h_r$ ratios ($r_{eff}/h_r=0.45\pm0.28$). Additionally, when we compare the structural parameters of table \ref{results}, with the results in \citet{labarbera_2010}, we find that \so\ lies below the Kormendy relation (either for $J$-- as for $K_s$--band), consistent with \citet{gadotti_2009} who find that pseudobulges do not tend to follow the Kormendy relation. Finally, if a galaxy hosts a pseudobulge, its center should be mostly population I material \citep[young stars, gas and dust,][]{kormendy_2013}. If we bear in mind that, in cases of high recent starburst, supergiants contribute to $K$--band luminosity \citep{minniti_1996}, then the $J-K_s$ color gradient of the host galaxy of \so, is in agreement with the latter pseudobulge classification criteria, where we see that, in the central region, $K_s$--band luminosity is stronger in comparison to $J$--band than in any other region of the galaxy.\smallskip

So far, only one galaxy able to launch a relativistic jet powerful enough to accelerate particles up to $\gamma-$ray energies, is known to host a pseudobulge; PKS 2004-447 \citep{kotilainen_2016}. Nevertheless, \citet{leontavares_2014} do not discard that the $\gamma-$\sy\ 1H 0323+342 is also hosted by a pseudobulge.\smallskip

We now evaluate whether the parameters derived for the bar in \so\ are in accordance with those for active early--type galaxies. Using the maximum ellipticity of the ellipse fits to the bar region as bar length \citep{marinova_2007}, we find that the length of the bar in \so\ is $r_{bar}=8.13\ \textrm{kpc}\pm0.25$, if we normalized it to the disc scale length $h_r$, we obtain $r_{bar}/h_r=1.00\pm0.06$. On the other hand, we can calculate the bar strength $f_{bar}$  \citep[][see also \citealt{whyte_2002,aguerri_2009,laurikainen_2007,hoyle_2011}]{abraham_merrifield_2000} defined as:
    \begin{center}
	\begin{equation}
	f_{bar}= \dfrac{2}{\pi}\Bigg\{arctan\Big[(b/a)^{-0.5}\Big]-arctan\Big[(b/a)^{0.5}\Big]\Bigg\}
	\label{eq:bar_strength}
	\end{equation}
	 \end{center}
\noindent
where b/a is the minor to major axis ratio of the bar. We obtain a bar strength $f_{bar}=0.17\pm0.03$. According to e.g. \citet{aguerri_2009} and \citet{laurikainen_2007}, the bar in \so\ is long and weak, consistent with S0 galaxies as found by \citep{laurikainen_2002}.\smallskip

The bar in \so\, might be related to the ring through resonances \citep[][and references therein]{athanassoula_2010}, given that secular evolution is likely the main evolutionary process that is currently in progress in its host galaxy. Therefore, the ring--like feature might be the result of gas redistribution by angular momentum transport driven by the bar (i.e. a ring constructed by a rotating bar interacting with the disk gas). In this scenario, the gas is moved by the bar into orbits near dynamical resonances \citep[for a review, see][]{athanasoula_2013}.\smallskip 

The main resonances are the Inner Lindblad Resonance (ILR) $\Omega_{ILR}=\Omega_{bar}-\kappa/2$, the Outer Lindblad Resonance (OLR) $\Omega_{OLR}=\Omega_{bar}+\kappa/2$ \citep{buta_combes_1996} and the Ultra Harmonic Resonance (UHR) $\Omega_{UHR}=\Omega_{bar}-\kappa/4$, where $\Omega_{bar}$ is the bar pattern speed and $\kappa$ is the epicyclic frequency  \citep{lindblad_1974}. The latter is located close to corotation \citep{sellwood_2013}, i.e. where the disc and the bar corotate. \smallskip 

Observations state that bars end near corotation 
\citep{kent_1987,sempere_1995,merrifield_kuijken_1995,gerssen_kuijken_merrifield_1999,debattista_2001,gerssen_2002,corsini_2002}, and can drive structures such as inner rings approximately at the UHR \citep{kormendy_2004}. Therefore, the feature located at $R=3.20''$, might be consistent with that of an inner ring. 


Another scenario for the ring formation in \so\ is a minor merger event. \citet{athanassoula_1997} show that the interaction of a small satellite galaxy on a barred galaxy can produce a ring that encloses the bar. Also, \citep{mapelli_2015} show that minor mergers with gas--rich satellites might explain the formation of rings in lenticular galaxies. This scenario is supported by the residuals in $J$--band (see Figure \ref{residuals_j}), where a feature of surface brightness $\mu=21.0\pm0.5\ \mathrm{mag}/\mathrm{arcsec}^2$ is shown about $\sim5.15''/13.10 kpc$ east from the center of \so\ \citep[resembling the Seyfert galaxy NGC 1097 whose light distribution is strongly affected by a small satellite galaxy,][]{higdon_2003}. This feature seems to interrupt the shape of the ring in the eastern part of the galaxy and even cause the color enhancement at $3.0''$. \smallskip 

An alternative and more likely scenario was proposed by \citet{marino_2011} for their sample of lenticular galaxies. The formation of the ring might be a joint effect of secular evolution driven by the bar and gas accreted from a small satellite galaxy (or many). Moreover, since S0 galaxies lack of an own gas reservoir (unlike spirals), this scenario also explains the origin of the gas needed to grow a massive bulge ($M_J=-22.42\pm0.40$ and $M_{Ks}= -24.21\pm0.32 $) and activate the black hole in \so, as well as the way this gas is channelled to the most central parts of the galaxy \citep[i.e. through angular momentum transport driven by the bar,][]{shlosman_1990,ohta_2007}.\smallskip

We finally observe that the parameters of the bar and the ring hosted by \so, are similar to the bar of PKS 2004-447 \citep{kotilainen_2016} and the ring in 1H 0323+342 \citep{leontavares_2014}. While the bars of PKS 2004--447 and \so, are $r_{bar}\approx7.80\textrm{ kpc}$ and $r_{bar}=8.13\pm0.25\textrm{ kpc}$ (taking the length of the bar as the maximum in the ellipticity profile), respectively, with absolute $K_s$--band magnitude  of $K_{bar}=-23.44\pm0.38$ and $K_{bar}=-23.86\pm0.52$, respectively; the rings of 1H 0323+342 and \so, are $\sim8.24\textrm{ kpc}$ and $\sim 8.13 \textrm{ kpc}$, respectively. Moreover, PKS 2004--447 shows an arm--like feature, whose origin might be related to a minor merger event \citep[see Figure 19 of][]{athanassoula_1997} and that, at some point, might become a ring, similar to the feature shown in \so.\smallskip

According to the most accepted processes for jet formation, the Blandford--Znajek \citep[BZ,][]{BZ_1977,BZ_1982,BZ_2013} and the Blandford--Payne \citep[BP,][]{BP_1982} mechanisms, the jet launching and collimation requires very massive black holes with high spins and strong magnetic fields. All of this require major mergers to occur, which fits well with previous observations \citep{mclure_2004, sikora_2007} and the jet formation paradigm \citep[where powerful relativistic jets are launched from giant elliptical galaxies,][]{marscher_2009}. However, it comes completely at odds with the morphology of \so, with a bar and a disc, that lacks of a classical bulge and with a black hole mass \citep[as estimated by the FWHM of its BLR lines and the continuum luminosity,][]{yuan_2008} $M_{BH}\sim8\times10^6M_\odot$ \citep[although, previous studies show that values $M_{BH}\gtrsim10^8M_{\odot}$ could be obtained when estimating its black hole mass by different methods,][]{baldi_2016, calderone_2013}.\smallskip


\section{Summary}
We have performed a detailed photometric analysis of the \g--\sy\ \so.\ We use deep near--infrared imagery in $J-$ and $K_s-$bands taken with the near--infrared camera NOTcam on the NOT. The main results of this analysis are:

\begin{itemize}
\item{The surface brightness distribution of \so\ is best fitted by a model resulting from a sum of a nuclear source, a bulge and a disc. Additionally to these components, a stellar bar in the $K_s$--band image is detected and modeled. The morphological parameters derived from our analysis show that the bulge, the disk and the bar of the host galaxy of \so\ fulfil the characteristics of SB0 galaxies.} 

\item {We find that the S\'ersic index and the relations between bulge and disc for \so\ are in good agreement with those of pseudobulges. Therefore, the bulge in the host galaxy of \so\ is statistically most likely to be pseudo.} 

\item {In both $J$-- and $K_s$--bands, we detect a ring enclosing the bar that is interrupted by, what it seems to be, a recent minor merger which might hint to the formation process of such inner ring, as suggested by \citet{athanassoula_1997}.}

\item {When comparing the ring and bar in \so\ to the ring and bar in 1H 0323+342 and PKS 2004-447 (the only two $\gamma-$\sy\ whose morphology have been analyised until now), we find similarities regarding size and magnitude. Likewise, PKS 2004--447 shows an arm--like feature, whose origin might be related to a minor merger event and that, at some point, might become a ring, similar to the inner ring in \so.}

\end{itemize}
We conclude that the prominent bar in the host galaxy of \so\ has mostly contributed to its overall morphology driving a strong secular evolution, which plays a crucial role in the onset of the nuclear activity and the growth of its massive (pseudo) bulge. Minor mergers, in conjunction, are likely to provide the necessary fresh supply of gas to the central regions of the host galaxy. \smallskip

Although our findings strongly suggest that secular evolution is the main process taking place in \so, our available data is insufficient to address some other questions as whether its (pseudo) bulge shows an increased star formation activity or if it is rotation--dominated \citep[as it should, given its disky origin;][]{kormendy_2013}. Therefore, we encourage different wavelengths imaging and integral field spectroscopy (IFS) observations to this galaxy and the whole sample of radio--loud \sy s.

\section*{Acknowledgements}

We thank Kari Nilsson who provided expertise that assisted this work.
We acknowledge support by CONACyT research grant 151494 (Mexico), CONACyT program for PhD studies and Finnish Centre for Astronomy with ESO (FINCA). 
This  research is  based on  observations made with the Nordic Optical Telescope, operated by the Nordic Optical Telescope Scientific Association at the Observatorio del Roque de los Muchachos, La Palma, Spain, of the Instituto de Astrof\'isica de Canarias. 
This publication makes use of data products from the Two Micron All Sky Survey.




\bibliographystyle{mnras}
\bibliography{ms_12dec15}







\bsp	
\label{lastpage}
\end{document}